\documentclass[11pt]{article}
\usepackage[margin=0.95in]{geometry}
\usepackage{epsfig,amssymb,soul}
\usepackage[normalem]{ulem}

\usepackage{hyperref}
\usepackage{graphicx}
\usepackage{epstopdf}
\usepackage{amsfonts,amsmath}
\usepackage{bm}
\usepackage{setspace}
\usepackage{comment}
\usepackage{siunitx}
\usepackage{mathrsfs}
\usepackage{multirow}
\usepackage{subcaption}
\usepackage[hang,flushmargin,bottom]{footmisc}
\usepackage{mathtools}
\usepackage{lipsum}
\usepackage{array}
\usepackage{enumerate}
\usepackage{ifthen}
\usepackage[affil-it]{authblk}
\usepackage{cite}
\usepackage{soul}
\usepackage{leftidx}
\usepackage{relsize}
\usepackage{braket}
\usepackage{float}
\usepackage{chngcntr}
\usepackage{mathrsfs}
\usepackage{bm}
\usepackage{cleveref}
\usepackage[left]{lineno}
\usepackage[dvipsnames]{xcolor}

\newcommand{\etal}{\textit{et al}.~}
\usepackage{comment}

\title{Analysis of nonlocal smart beams following fractional-order constitutive relations}

\author[1]{Shubham Desai}
\author[1]{Sai Sidhardh\footnote{\textit{Corresponding author: sidhardh@mae.iith.ac.in}}}
\affil[1]{Department of Mechanical and Aerospace Engineering, IIT Hyderabad, Kandi, Telangana 502285, India}

\date{}

\begin{document}
	
\maketitle
\begin{abstract}
In this study, we develop a fractional-calculus based constitutive model for capturing nonlocal interactions over the multiphysics response in solids. More specifically, we develop constitutive relations for nonlocal piezoelectricity incorporating fractional-order kinematic relations to capture the long-range interactions over electrical and mechanical field variables. This study breaks new ground by developing fractional-order constitutive models for a two-way multiphysics (electro-mechanical) coupling, specifically the direct and converse piezoelectric effect. It is expected that long-range interactions over each field variable (elastic and electrical) can be leveraged to develop metastructures with enhanced multiphysics coupling. To better illustrate this, we choose the example of a smart beam composed of a nonlocal substrate and a piezoelectric layer. We establish the analytical and numerical framework to analyze nonlocal smart beams based on variational principles. The fractional-Finite Element (f-FE) numerical solver, facilitating multiphysics coupling, undergoes comprehensive validation through multiple case studies. Finally, detailed studies point towards tuning the multiphysics coupling possible via nonlocal interactions across the domain.
\end{abstract}

\section{Introduction}
It is becoming increasingly possible to fabricate complex and intrinsically multiscale structures having intricate material distributions and geometries. Examples include sandwich structures, metallic foams, porous materials, periodic structures, and composites. Clearly, significant improvements in elastic behavior can be demonstrated by tailoring the microstructure of such complex structures \cite{wu2018effective,wen2022perspective}.  It is established that such complex structures can be realized as nonlocal structures with long-range interactions across the domain that influence the response at a point due to either material or geometric heterogeneities. For instance, the response of porous media \cite{patnaik2022role}, sandwich composites\cite{romanoff2021potential}, or even biological materials like tissues\cite{vadgama20052} exhibits experimentally observed scale-dependent effects. Such nonlocal interactions are not limited to elastic response but have been noted in other physical fields like electrodynamics\cite{tarasov2022general,kosztin1997nonlocal,hildebrandt2004novel,kornyshev1978model,valentini1991non}. 

Smart materials offer the advantage of exploiting multiphysics coupling towards achieving multiple, sometimes including structural, objectives. However, their application has limitations owing to typically weak coupling exhibited by natural materials. For instance, the piezoelectric effect, an electro-mechanical coupling of the mechanical strain to the electrical field, is rather weak, as evidenced by the very low coupling coefficients\cite{anand2020continuum}. Further, energy transfer between electrical and mechanical fields in these materials is effective at specific (resonant) loading conditions\cite{dietl2010timoshenko,zhang2007stability,giri2021dynamics}.  Although several approaches to enhance the multiphysics coupling have been explored, limited research exists on exploiting the scale effects in macrostructures for enhancing the multiphysics coupling. 
Of interest are works exploring nonlocal interactions for electrodynamics\cite{tarasov2022general}, electro-magnetism\cite{tarasov2008fractional,tarasov2009fractional}, among others. This may be further expanded to develop metastructures with enhanced multiphysics coupling for application in wide-ranging domains. 

The classical integer-order constitutive models are not compatible with the scale-dependent phenomenon\cite{fleck1997strain,romanoff2021potential}. To address this gap, gradient elasticity theories\cite{mindlin1968first,toupin1962elastic,eringen1964nonlinear}, and integral elasticity theories \cite{eringen1983differential,romano2017constitutive} have been proposed to capture the scale-effects. These higher-order constitutive theories employ different approaches to capture the long-range effects at a point in the solid. Eringen extended the integral theory for nonlocal elasticity to model long-range interactions over the electro-mechanical response of piezoelectric materials \cite{eringen1984theory}. However, it is well documented that these integer-order constitutive models for nonlocal interactions suffer from limitations. For instance, the physical significance of the additional boundary conditions corresponding to strain-gradient terms is still unclear and is an active area of research \cite{peerlings2001critical,aifantis2003update,sidhardh2019exact}. In contrast, integral methods are afflicted by mathematical ill-posedness, resulting in imprecise predictions, including the failure to account for nonlocal effects in some cases or the emergence of undesired hardening behavior \cite{romano2017constitutive,sidhardh2021thermodynamics}. It must be noted that these inconsistencies are fundamental, resulting from a violation of the thermodynamic laws\cite{sidhardh2021thermodynamics,polizzotto2001nonlocal}.

There exists a sizable literature on nonlocal piezoelectricity following the differential gradient theories discussed above for nonlocal interactions. Employing these constitutive theories, Liu \etal \cite{liu2013thermo} developed analytical models to analyze nonlocal effects on the piezoelectric plate. While increasing the degree of nonlocal interaction, the consistent softening of the elastic response has been observed in the study. However, similar analyses on the effect of long-range interactions over the piezoelectric coupling have not been studied. Similarly, Arefi \cite{arefi2018analysis} demonstrated a consistent softening of the mechanical response of piezoelectric structures due to long-range interactions, but the influence on electro-mechanical coupling was inconclusive. Liu \etal \cite{liu2021reflection} studied the elastodynamic response of nonlocal piezoelectric structures and realized the effect of nonlocal interactions over mechanical and electrical field distributions towards enhancement of the piezoelectric coupling. Liao \etal \cite{ke2015free} noted an unexpected lack of nonlocal effects on the free vibration behavior of nonlocal piezoelectric plates under particular boundary conditions. Recently, Naderi \etal \cite{naderi2021local} established the incompetence of differential models for capturing long-range interactions in the analysis of nonlocal piezoelectricity. It must be noted that in most of the existing literature, the influence of nonlocal effect has been observed only on the elastic response of the piezoelectric solid.

In recent years, fractional calculus has witnessed significant developments across various fields as an effective mathematical framework for modeling diverse nonlocal and multi-scale phenomena. Fractional derivatives, a class of operators that encompass both differentiation and integration, possess inherent multi-scale properties and offer a convenient approach to incorporate nonlocal effects into mathematical models\cite{cottone2009elastic, di2008long, carpinteri2014nonlocal, sumelka2014fractional, sumelka2015fractional}. The extensive utilization of fractional calculus in nonlocal elasticity is attributed to this inherent multi-scale characteristics of fractional operators. This approach has proven effective in addressing the limitations of classical methods by employing fractional-order kinematic relations within a nonlocal continuum framework\cite{patnaik2020ritz, sidhardh2021thermodynamics, patnaik2021towards}. Furthermore, using a nonlocal framework with fractional-order kinematic interactions yields well-posed governing equations and, as a result, unique solutions. Additionally, the efficacy of fractional-order continuum theories in developing reduced-order models for complex heterogeneous systems is already established\cite{hollkamp2019analysis,hollkamp2020application,patnaik2020generalized,patnaik2022fractional}. 
Further, this area of research is evolving and has potential for significant breakthroughs. Fractional calculus holds significant potential for exploring nonlocal effects in electrostatics and electrodynamics attributed to spatially dispersive permittivities realized at lower length scales\cite{hildebrandt2004novel}, interaction with environment, memory and distributed lag (temporal nonlocality. The concept of nonlocal interactions in these fields was introduced nearly fifty years ago by Kornyshev et al. \cite{kornyshev1978model}. Similar to nonlocal elasticity, various integral and differential approaches have been investigated to model long-range interactions \cite{hildebrandt2004novel,kornyshev1978model,valentini1991non}. Recently, fractional-order derivatives have also been applied to the governing equations in electrical systems by Tarasov \cite{tarasov2022general}. 

Limited research exists on the application of fractional-order derivatives in developing constitutive models for multiphysics problems\cite{sidhardh2021thermodynamics,tarasov2019handbook}. Sidhardh \etal developed analytical and numerical models for fractional-order constitutive theories on thermoelasticity\cite{sidhardh2021thermodynamics,patnaik2022role}. More specifically, the nonlocal effects on the thermoelastic response of complex structures are captured via a fractional-order derivative definition for mechanical strain. However, this is a one-way coupling focusing on the influence of thermal effects over nonlocal elastic response, but the effect of nonlocal elastic response over the temperature distribution is not considered. Clearly, literature on fractional-order constitutive models for two-way multiphysics coupling is limited. Additionally, a quantitative analysis of the nonlocal effects on the multiphysics coupling requires to be explored. Given the prevalence of applications involving piezoelectric materials, we propose developing an analytical framework based on fractional calculus for nonlocal effects on piezoelectricity. This is in net contrast to alternate formulations for nonlocal piezoelectricity employing integer-order constitutive relations.

Most literature on fractional-order models in elasticity focuses on developing analytical models, with limited numerical studies to better illustrate the quantitative effect of nonlocal interactions. This is due to difficulties developing numerical models suitable for solving fractional-order differential equations. Although several finite element formulations for integer-order equations have been previously extended to fractional-order equations, they are mostly restricted to temporal fractional-order derivatives that provide hyperbolic or parabolic differential equations \cite{agrawal2008general, deng2009finite, zheng2010note,liu2015two}. Patnaik \etal \cite{patnaik2020ritz,sidhardh2020geometrically} proposed and developed a finite element solver for general spatial fractional-order governing equations in recent years. More recently, Rajan \etal \cite{rajan2024element} developed a mesh-free solver based on an element-free Galerkin method for the numerical solution of fractional-order differential equations. These solvers are all restricted to a numerical solution of single (elastic) field variables. So, a numerical model is proposed to be developed here to illustrate the influence of nonlocal effects over multiple field variables. 

A challenge in realization of fractional-order constitutive models for for nonlocal elasticity is the development of a physically unique method of characterization for additional constitutive parameters. It is documented in literature that there are two ways to determine the fractional-order constitutive parameters for a complex structure: using physics-based methods, and data-driven approach. The former was employed in the characterization of fractional-order homogeneous elastic model equivalent of: (i) a grid-stiffened plate\cite{patnaik2020generalized}, (ii) periodic structures \cite{patnaik2022fractional}, and  (iii) a beam with varying cross section\cite{hollkamp2020application}. Such physics-based approaches rely either on matching the deformation energy\cite{karttunen2020hierarchy}, or any other similar parameter relevant for the physics of the problem being studied\cite{patnaik2020generalized,patnaik2022fractional}. More recently, data-driven approaches based on inverse problems for material characterization are also being established for an estimation of the fractional-order constitutive parameters for nonlocal elasticity\cite{ding2022multiscale,patnaik2022variable}.

In this paper, we start with a review of the fractional-order constitutive models for nonlocal solids. Following this, a detailed derivation of the constitutive relations for nonlocal piezoelectricity, based on fractional-order kinematic relations for electrical field and mechanical strain, is provided in \S~\ref{sec: constt.}. Thereafter, considering the example of a slender smart beam, the fractional-order constitutive relations proposed here are employed to develop the integro-differential governing equations of equilibrium for direct and converse piezoelectric coupling in \S~\ref{sec:problem_formulation}. Subsequently, a numerical tool is developed in \S~\ref{sec:FEM} for the solution of fractional-order multiphysics differential equations. Finally, a detailed analysis is reported in \S~\ref{sec: results} on the effect of the long-range interactions over the direct and converse piezoelectric coupling, including a discussion on opportunities to leverage the scale-effects for tuning the electro-mechanical coupling. 
\section{Constitutive model for fractional-order piezoelectricity}
\label{sec: constt.}
In this section, the fractional-order constitutive model for nonlocal piezoelectricity will be developed. 
Following a fractional-order definition for the mechanical strain and the electric field variables, the strain-displacement (kinematic) and stress-strain (material) constitutive relations for nonlocal piezoelectric solid will be re-defined. Finally, energy norms for a fractional-order piezoelectric solid will be provided here for developing governing differential equations of equilibrium/motion.
	
\subsection{Fractional-order nonlocal continuum modeling}
\label{subsec: frac_cont}
We briefly review the previously developed nonlocal model for elastic beam via fractional-order approach in \cite{patnaik2020generalized,sidhardh2021thermodynamics}. In the current context of nonlocal piezoelectricity, these definitions will be extended to electrical field variables. 
	
Following a displacement-driven approach to modeling nonlocal interactions, the strain-displacement relations are recast employing fractional-order derivatives as follows\cite{sidhardh2020geometrically}:
\begin{equation}
	\label{eq:strain}
        \tilde{\bm{\epsilon}} = \frac{1}{2}{(\mathbf{D}^{\alpha_m}_\mathbf{X} \mathbf{U} + (\mathbf{D}^{\alpha_m}_{\mathbf{X}} \mathbf{U})^T)}
\end{equation}
where $\mathbf{U}(\mathbf{X})$ is the displacement field at any point $\mathbf{X}$ in the nonlocal solid, and $\mathbf{D}^{\alpha_m}_\mathbf{X} \mathbf{U} $ is the fractional-order displacement gradient defined as \cite{sidhardh2021thermodynamics,sidhardh2020geometrically}:
\begin{equation}
    \begin{aligned}
	\label{eq:RC_def}
        \mathbf{D}^{\alpha_m}_\mathbf{X} \mathbf{U} = \frac{1}{2}\Gamma(2-{\alpha_m})\left[{l_A^{{\alpha_m}-1}}{\left({{{_{\mathbf{X}_A}^C}}D{_\mathbf{X}^{\alpha_m}}}\mathbf{U}(\mathbf{X})\right)}-{{l_B^{{\alpha_m}-1}}}{\left({{_\mathbf{X}^C}D\mathbf{_{X_B}^{\alpha_m}}}\mathbf{U}(\mathbf{X})\right)}\right]
    \end{aligned}
\end{equation}
The above defined space fractional derivative $\mathbf{D}^{\alpha_m}_\mathbf{X} \mathbf{U}$ follows a Riesz-Caputo (RC) definition with the fractional-order ${\alpha_m} \in (0,1]$ and it is defined on the interval ${\mathbf{X} \in (\mathbf{X_A},\mathbf{X_B})}$; ${{{_{\mathbf{X}_A}^C}}D{_\mathbf{X}^{\alpha_m}}\mathbf{U}(\mathbf{X})}$ is the left Caputo fractional-order derivative of order $\alpha_m$ over the domain $(\mathbf{X_A},\mathbf{X})$ and ${{{_\mathbf{X}^C}D\mathbf{_{X_B}^{\alpha_m}}}\mathbf{U}(\mathbf{X})}$ is the right Caputo fractional-order derivative of the same order over the domain $(\mathbf{X},\mathbf{X_B})$. Further, we define ${l_A=\mathbf{X}-\mathbf{X}_A}$ and ${l_B=\mathbf{X}_B-\mathbf{X}}$ are the nonlocal horizon of influence to the left- and right- of $\mathbf{X}$. The gamma function $\Gamma(.)$ and ${{l_A^{{\alpha_m}-1}}}$ and ${l_B^{{\alpha_m}-1}}$
are considered in the above definition to ensure the fractional-order definition for strain is frame-invariant and dimensionless. Complete details regarding the development of the above expression are available in \cite{patnaik2020generalized,patnaik2021towards}. 
Finally, the domain $(\mathbf{X_A}, \mathbf{X_B})$ referred to here as the horizon of nonlocality at $\mathbf{X}$ is defined by the interval of the fractional-order derivative. This is analogous to the attenuation range in the literature on classical nonlocal elasticity\cite{eringen1972nonlocal}.
	
Extending the above formalism to multiphysics variables, we propose to develop a fractional-order constitutive model for nonlocal electrostatics. 
Analogous to a displacement-driven approach to nonlocal elasticity, possible thanks to fractional-order constitutive relations, nonlocality is introduced in the kinematic constitutive relations of electrostatics \cite{tarasov2019handbook}. For this purpose, the fractional-order definition of the electric field is given as:
\begin{equation}
    \label{eq:non_local_ele}
    \mathbf{\Tilde{E}} = -\mathbf{D}^{\alpha_e}_{\textbf{X}}{{\phi}}
\end{equation}
where, $\mathbf{D}^{\alpha_e}_{\textbf{X}}[\cdot]$ follows the RC definition given in Eq.~\eqref{eq:RC_def}. Here $ \mathbf{\Tilde{E}}$ is the nonlocal electrical field, ${\phi(\textbf{X})}$ is the scalar electrical potential field, and $\alpha_e$ is the fractional-order capturing the nonlocal interactions over the electrical field. 
As mentioned earlier in Eq.~\eqref{eq:RC_def}, RC definition for fractional-order derivative alleviates the issues such as frame invariance, objectivity, and dimensional consistency with alternate fractional-calculus based models for nonlocal electrostatics. Note that in the integer-order model (classical model), the electric field at a point depends only on the electrical potential in its immediate surroundings. In contrast, the fractional-order model defined above captures the nonlocal interactions over the electric field evaluated at the point of interest. More clearly, the electric field at a point $\textbf{X}\in \Omega$ depends not only on the electric potential at this point but also on the electric potential at points within a horizon of nonlocal influence. The long-range interactions are captured by the differ-integral nature of the fractional-order derivatives employed in the above equation. 
	
In the above, the influence of long-range interactions on mechanical field variables is captured by order $\alpha_m$, and on electrical field variables is captured by order $\alpha_e$. For a general dielectric material subject to arbitrary electro-mechanical loads, $\alpha_m$ and $\alpha_e$ are independent parameters. The constitutive parameters introduced via FC models for elastic and electrical field variables, while analogous, are independent constants and may differ from each other. 

\subsection{Constitutive relations for nonlocal piezoelectricity}
\label{subsec: constt.}
In this section, the material constitutive relations for a nonlocal piezoelectric solid will be developed following the fractional-order kinematics presented in \S~\ref{subsec: frac_cont}. More clearly, the Helmholtz free energy for the nonlocal piezoelectric solid will be proposed, and the stress-strain relations for the nonlocal piezoelectric solid will be derived.
	
The Helmholtz free energy density for the nonlocal piezoelectric solid, following the fractional-order constitutive relations developed above, is defined as follows:
\begin{equation}
    \label{eq: helm_np}
    \mathcal{H}(\tilde{\bm{\epsilon}},\tilde{\mathbf{E}})=-\underbrace{\frac{1}{2}\tilde{\mathbf{E}}\cdot\mathbf{a}\cdot\tilde{\mathbf{E}}}_{\text{Electrical energy}}+\underbrace{\frac{1}{2}\tilde{\bm{\epsilon}}: \bm{C}:\tilde{\bm{\epsilon}}}_{\text{Mechanical energy}}- \underbrace{\tilde{\mathbf{E}} \cdot \mathbf{e}: \tilde{\bm{\epsilon}}}_{\text{Piezoelectric coupling energy}}
\end{equation}
where $\mathbf{a}$ is the second-order dielectric permittivity, $\bm{C}$ is the fourth-order elasticity constant, and $\mathbf{e}$ is the third-order piezoelectric coefficient. The fractional-order strain and electric field in the above expression are defined in Eqs.~\eqref{eq:strain} and \eqref{eq:non_local_ele}, respectively.
The expression above corresponds to the free energy density of a nonlocal piezoelectric solid incorporating long-range interactions over both the mechanical and electrical field variables. It is of interest to note that the energy term corresponding to piezoelectric coupling is influenced by the nonlocal interactions over mechanical and electrical field variables. The above expression is defined in a manner analogous to local piezoelectricity\cite{anand2020continuum}. While the above expression may also be derived rigorously, it is skipped here for the sake of brevity. Interested readers may refer to \cite{sidhardh2021thermodynamics} for a detailed derivation of energy in terms of independent field variables as a Taylor series expansion.
	
The stress in the nonlocal piezoelectric solid can now be given as follows\cite{anand2020continuum,sidhardh2021thermodynamics}:
\begin{equation}
    \label{eq: constt_nl_piezo}
    \tilde{\sigma}_{ij}=\frac{\partial \mathcal{H}}{\partial \tilde{\epsilon_{ij}}}=C_{ijkl}\tilde{\epsilon}_{kl}-e_{kij}\tilde{E}_k,~~~~\tilde{D}_i=-\frac{\partial \mathcal{H}}{\partial \tilde{E_i}}=a_{ij}\tilde{E}_{j}+e_{ijk}\tilde{\epsilon}_{jk}
\end{equation}
Note that piezoelectric coefficient tensor $e_{ijk}$ provides the electro-mechanical coupling within the dielectric. In the above, $\tilde{\sigma}_{ij}$ and $\tilde{D}_{i}$ are the mechanical stress and electrical displacement at a point where the nonlocal strain $\tilde{\epsilon}_{kl}$ and the electric field $\tilde{E}_{i}$ are evaluated as given in Eqs.~\eqref{eq:strain} and \eqref{eq:non_local_ele}. The above constitutive relations show a point-to-point correspondence between mechanical stress and electrical displacement to mechanical strain and electrical field variables. However, it must be clarified that the stress and electrical displacement are nonlocal and include the long-range interactions via fractional-order definitions for strain and electric field. Recall that point-to-point correspondence between mechanical stress-strain (and electrical field displacement) is essential for the thermodynamic consistency of the constitutive relations for nonlocal continuum as demonstrated in \cite{sidhardh2021thermodynamics}.
	
Finally, using the constitutive relations developed in Eq.~\eqref{eq: constt_nl_piezo}, the Helmholtz free energy can be recast as follows:
\begin{equation}
    \label{eq: helm_np2}
    \mathcal{H}(\tilde{\bm{\epsilon}},\tilde{\mathbf{E}})=\frac{1}{2}\bm{\tilde{\sigma}}:\bm{\tilde{\epsilon}}-\frac{1}{2}\tilde{\mathbf{D}}:\tilde{\mathbf{E}}
\end{equation}
It must be pointed out that the long-range interactions over the mechanical and electric fields are realized independently by the fractional-order strain and electric field, respectively. However, on account of non-zero electro-mechanical coupling, the stress and electrical displacement simultaneously realize long-range interactions over both the mechanical and electric fields. Finally, as expected, for a choice of $\alpha_m=\alpha_e=1$, the constitutive relations for the local piezoelectric solid are recovered.

\section{Problem formulation}
\label{sec:problem_formulation}
\begin{figure} [H]
    \centering
    \includegraphics[width=0.75\linewidth]{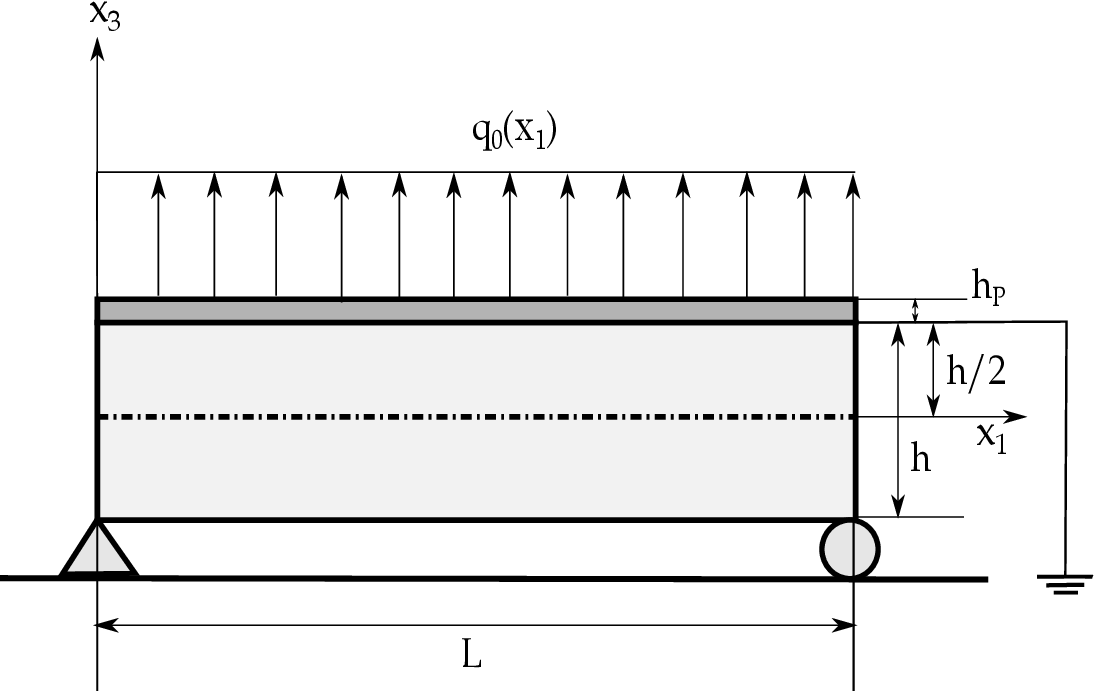}
    \caption{Schematic diagram of unimorph piezoelectric smart beam subject to a uniformly distributed transverse load $q_0(x_1)$.}
    \label{fig:Eular_Bernaulli_Beam}
\end{figure}

A schematic diagram of a simply supported smart beam is presented in Fig.~\ref{fig:Eular_Bernaulli_Beam}. The smart beam comprises an elastic substrate and a piezoelectric patch (varying in length and location) attached to the top surface of the elastic substrate (unimorph configuration).
As shown in Fig~\ref{fig:Eular_Bernaulli_Beam}, the Cartesian reference frame adopted in this study, the positions of the ends of the smart beam in the ${x_1}$-direction at ${x_1}=0$ and ${x_1}=L$. The ${x_3}$-axis is aligned such that ${x_3}=0$ corresponds to the mid-plane of the elastic substrate beam. Additionally, the bottom surface of the elastic beam is situated at ${x_3}=-h/2$, while the top surface of the elastic substrate, including a piezoelectric patch with a thickness of $h_p$, is located at ${x_3}=h/2$.
	
In this study, we consider the beam to be slender ($L/h>50$) and proceed with an Euler-Bernoulli beam displacement theory to develop the analytical and numerical model for the elastic response of the smart beam. The axial displacement $u_1({x_1},{x_3})$, and the transverse displacement $u_3({x_1}, {x_3})$ at any point within the beam are given as:
\begin{equation}
    \label{eq:Disp_field}
    {u_1}(x_1, x_3)={u_0}({x_1}) - {x_3}\left[\frac{\mathrm{d}{w_0}({x_1})}{\mathrm{d}{x_1}}\right],~~~~~~~~~~~{u_3}({x_1, x_3})={w_0}({x_1})
\end{equation}
where ${u_0}({x_1})$ and ${w_0}({x_1})$ are the mid-plane axial and transverse displacements ($x_3=0$). 
For the above-given displacement field, the axial strain at any point in the smart beam evaluated following Eq.~\eqref{eq:strain} is given as:
\begin{equation}
    \label{eq:kinematics}
    {\Tilde{\epsilon}_{11}}({x_1}, {x_3})={D_{{x}_1}^{\alpha_m}}{u_0}({x_1})-{x_3}{D_{{x}_1}^{\alpha_m}}\left[\frac{\mathrm{d}{w_0}({x_1})}{\mathrm{d}{x_1}}\right]
\end{equation}
Recall from \cite{patnaik2020ritz} that these length scales are position-dependent and defined such that the nonlocal horizon of influence for points close to boundaries/discontinuities is appropriately truncated at these boundaries. Therefore, these nonlocal length scales in the substrate and the piezoelectric patch are independent variables. To differentiate the fractional-order derivatives over the substrate and piezoelectric layer, we use the superscript $(\cdot)^P$ for variables corresponding to the piezoelectric material. Also, the fractional-order $\alpha_m \in (0,1]$. In the above, the transverse shear strain is neglected. The non-zero transverse shear strain (possible following Eq.~\eqref{eq:strain}) is neglected here on account of the beam being assumed to be slender with $L/h>100$. In these cases, the rigidity against transverse shear force significantly surpasses the rigidity against the bending ($K_{shear}/K_{bend}\propto(L/h)^2$) \cite{reddy2019introduction}. Consequently, the shear deformation of the slender beam can be neglected.
The electrical potential $\phi(x_1,x_3)$ is assumed to vary linearly across the thickness of the slender structure \cite{junior2009electromechanical}. Therefore, the electrical field across the piezoelectric layer is defined as:
\begin{equation}
    \label{eq: elec_potential}
    \phi(x_1,x_3)=\phi_0(x_1)\frac{(x_3-h/2)}{h_P}
\end{equation}
where the bottom surface of the piezoelectric patch ($x_3=h/2$) is grounded as indicated in the schematic in Fig.~\ref{fig:Eular_Bernaulli_Beam}, and the electrical potential at the top surface ($x_3=h/2+h_P$) is $\phi_0(x_1)$. Thereby, following Eq.~\eqref{eq:non_local_ele}, the fractional-order electrical fields corresponding to the above given electrical potential are:
\begin{equation}
    \label{eq:elec_fields}
    \tilde{E}_1(x_1,x_3)=-(D_{x_1}^{\alpha_e}\phi_0(x_1))~\frac{(x_3-h/2)}{h_P},~~~~~\tilde{E}_3(x_1)=-\frac{\phi_0(x_1)}{h_P}
\end{equation}
here $\alpha_e\in(0,1]$ is the fractional-order corresponding to nonlocal interactions between electric field variables over the solid domain. Note that the transverse electric field is local; it depends only on the electrical potential at the point of interest. This is because the fractional-order derivative of the potential along $x_3-$direction defined in Eq.~\eqref{eq: elec_potential} reduces trivially to integer-order derivatives. Physically, this corresponds to limited nonlocal interactions across the thickness within slender structures. Note that the remaining components of the electrical field and mechanical strain are zero for the displacement field and electrical potential chosen in Eqs.~\eqref{eq:Disp_field} and \eqref{eq: elec_potential}. We note here that, electric field along the $x_1-$direction ($\tilde{E}_1$) is non-zero for general distributions. This is particularly true for the open circuit configuration (studied later) that induces a non-uniform distribution for electrical potential at the top surface for an externally applied mechanical load via direct piezoelectric effect. However, the effect of non-zero $\tilde{E}_1$ is extremely weak (at least a couple of orders weaker than $\tilde{E}_3$: $E_1/E_3\propto{h_P/L_P}$), and hence may be neglected.

In the current study with a uniformly distributed electrical field, we can prescribe a single value for voltage at the Parallel Plate Electrode (PPE) provided on the top surface of the piezoelectric patch while connecting a similar PPE attached at the bottom surface to ground. A varying axial distribution for electric potential requires a more sophisticated configuration of piezoelectric elements with multiple instances of a set of  PPE along the length of the piezoelectric layer. A spatial distribution of the electrical potential over the entire substrate may be achieved by providing different externally applied voltages on each instance of piezoelectric element \cite{marinkovic2009aspects}. However in all subsequent studies, the effect of electrode on electro-mechanical response of a smart beam is assumed to be negligible. This is analogous to similar assumptions employed in several analytical and numerical studies \cite{}. A short note regarding the effect will be discussed later in \S~\ref{sec: results}
	
The material constitutive relations in Eq.~\eqref{eq: constt_nl_piezo} may now be used to give the mechanical stress and electrical displacement of the nonlocal piezoelectric solid as:
\begin{equation}
    \label{eq:piezo_mat_cons}
    \tilde{\sigma}_{11}^P(x_1, x_3)=E_P \tilde{\epsilon}_{11}^P(x_1, x_3)-e_{31} \tilde{E}_{3}(x_1),~~~\tilde{D}_{3}(x_1, x_3)=a_{33}\tilde{E}_{3}(x_1)+e_{31}\tilde{\epsilon}_{11}^P(x_1, x_3)
\end{equation}
where $E_P$ is the elastic modulus of the piezoelectric patch, $a_{33}$ is the dielectric permittivity, and $e_{31}$ is the piezoelectric constant. $\tilde{\epsilon}_{11}^P$ is the fractional-order strain in the piezoelectric solid evaluated following the RC definition in Eq.~\eqref{eq:RC_def} using the corresponding nonlocal length scales. Finally, as mentioned above, the classical (local) piezoelectric constitutive relations are exactly recovered for $\alpha_m=\alpha_e=1$.
	
Considering the substrate as a purely elastic nonlocal structure, the corresponding material constitutive relations are given as follows:
\begin{equation}
    \label{eq:subs_const_relations}
    \tilde{\sigma}_{11}(x_1, x_3)=E_S~\tilde{\epsilon}_{11}(x_1, x_3)
\end{equation}
where $E_S$ is the elastic modulus of the substrate, it must be mentioned that the strain in the above equation ($\tilde{\epsilon}_{11}(x_1)$) follows from Eq.~\eqref{eq:kinematics} evaluated using nonlocal length scales for the substrate. 
Finally, note that the constitutive relations of classical (local) elasticity are exactly recovered for fractional-order $\alpha_m=1$. The functional dependence of the variables is skipped in the expressions below for the sake of convenience.

The deformation energy of the substrate is given by\cite{patnaik2020ritz}:
\begin{equation}
    \label{eq:energy_subs}
    \mathcal{U}_S=\int_{0}^{L}\int_{0}^{b}\int_{-h/2}^{h/2} \frac{1}{2}\tilde{\sigma}_{11}\tilde{\epsilon}_{11} ~\mathrm{d}x_3~\mathrm{d}x_2~\mathrm{d}x_1
\end{equation}

The agumented Helmholtz free energy for the piezoelectric patch, defined in Eq.~\eqref{eq: helm_np2} including nonlocal interactions, is given by:
\begin{equation}
    \label{eq:energy_piez}
    \mathcal{H}_P=\int_{x_0}^{x_0+L_P}\int_{0}^{b}\int_{h/2}^{h_P+h/2} \frac{1}{2}\left(\tilde{\sigma}_{11}^P\tilde{\epsilon}_{11}^P-\tilde{D}_{3}\tilde{E}_{3}\right) ~\mathrm{d}x_3~\mathrm{d}x_2~\mathrm{d}x_1
\end{equation}
here, $L_P$ is the length of the piezoelectric patch, and $x_0$ is the initial coordinate for the piezoelectric patch; $x_0=0$ for the piezoelectric patch starting at the left end of the smart beam. The internal energy $\mathcal{H}$ of the nonlocal elastic smart beam is:
\begin{equation}
    \label{eq:deforma_energy}
    \mathcal{H}=\mathcal{U}_S+\mathcal{H}_P
\end{equation}
Now, we derive the governing differential equations of the nonlocal smart beam following the fundamental principle of variational calculus \cite{reddy2019introduction}. The total energy of the nonlocal smart beam is given as:
\begin{equation}
    \label{eq:poten_eng}
    \bm{\Pi}=\mathcal{H}-\int_{0}^{L}{{f_a}{u_0}}\,\mathrm{d}{x_1}\ - \int_{0}^{L}{{f_t}{w_0}}\,\mathrm{d}{x_1}\--\int_{x_0}^{x_0+L_P}{{f_\phi}{\phi_0}}\,\mathrm{d}{x_1}\
\end{equation}
where, $f_a$ and $f_t$ are the external mechanical forces applied along the $x_1-$ and $x_3-$ directions, respectively and $f_\phi$ is externally applied electrical charge over piezoelectric material along $x_1-$direction.
	
Employing the principle of minimum potential energy over  Eq.~\eqref{eq:poten_eng} gives:
\begin{equation}
    \begin{aligned}
    \label{eq:variation}
    \delta{\bm{\Pi}} = & \int_{V}{\tilde{\sigma}_{11}}\delta{\tilde{\epsilon}_{11}}\mathrm{d}V + \int_{V_P}\left({\tilde{\sigma}_{11}^P}\delta{\tilde{\epsilon}_{11}^P} -\tilde{D}_{3}\delta\tilde{E}_{3}\right)\mathrm{d}V_P  - \int_{L}{{f_a}\delta{u_0}}\,\mathrm{d}{x_1}\ - \int_{L}{{f_t}\delta{w_0}}\,\mathrm{d}{x_1}\ -\int_{L_P}{{f_t}\delta{\phi_0}}\,\mathrm{d}{x_1}\
    \end{aligned}
\end{equation}
where $V$ and $V_P$ denote the volume of the substrate and the piezoelectric layer, respectively. The parameter $f_\phi$ is identically equal to zero for all studies on open circuit demostrating direct piezoelectric effect, on account of total charge vanishing at the electrode\cite{ray1993exact}. Further, in studies for converse piezoelectricity, where a prescribed electric potential is applied, this parameter does not appear in the governing equations as $\delta\phi=0$. Therefore, we omit this parameter altogether in all subsequent equations.
Following fractional-order strain-displacement relation in Eq.~\eqref{eq:kinematics} and electrical field definition in Eq.~\eqref{eq:elec_fields}, the above equation can be simplified as:
\begin{equation}
    \begin{aligned}
    \label{eq:simplified_variation}
    \delta{\bm{\Pi}} = &\int_{V}{\tilde{\sigma}_{11}}\left(D^{{\alpha_m}}_{x_1}\delta{u_0}-{x_3}D^{{\alpha_m}}_{x_1}\frac{\mathrm{d}\delta{w_0}}{\mathrm{d}{x_1}} \right)\mathrm{d}V + \int_{V_P}\left[{\tilde{\sigma}_{11}^P}\left(D^{{\alpha_m}}_{x_1}\delta{u_0}-{x_3}D^{{\alpha_m}}_{x_1}\frac{\mathrm{d}\delta{w_0}}{\mathrm{d}{x_1}} \right)\right.\\& \left. +\tilde{D}_{3}\frac{\delta\phi_0}{h_P}\right]\mathrm{d}V_P -\int_{L}{{f_a}}\delta{u_0}\,\mathrm{d}{x_1}\ - \int_{L}{{f_t}}\delta{w_0}\,\mathrm{d}{x_1}\
    \end{aligned}
\end{equation}
Finally, the integro-differential governing equations of equilibrium will be derived for the nonlocal smart beam via appropriate mathematical operations. This includes performing closed-form integration across the cross-section of the beam (substrate and piezoelectric) and integration by-parts for isolating the variation in independent field variables. These steps, although for a nonlocal purely elastic beam, are available in the literature\cite{patnaik2020ritz}. So, these steps are not presented in detail here for the sake of brevity. We directly proceed to the electrical and mechanical governing equations of equilibrium for the smart beam.
	
\subsection{Governing differential equations for converse piezoelectric effect}
\label{subsec:GDE_converse}
We begin with the governing differential equations for a nonlocal smart beam, where the piezoelectric layer is subject to electrical loading. More clearly, the nonlocal piezoelectric layer acts as an actuator, presenting a mechanical response to externally applied electrical potential. Therefore, in this case, the electrical potential in Eq.~\eqref{eq:elec_fields} is prescribed ($\delta \phi_0 = 0$). 
	
The governing differential equations of equilibrium of the smart beam ($x_1~\in~(0, L)$) with nonlocal interactions when subject to electro-mechanical loads are given as follows:
\begin{equation}
    \label{eq:gov_diff_eq_converse}
    \mathcal{D}^{\alpha_m}N(x_1)+\mathcal{D}_p^{\alpha_m}N^P(x_1)+f_a(x_1)=0
 ~~~~~
    \frac{\mathrm{d}}{\mathrm{d}x_1}\left[\mathcal{D}^{\alpha_m}M(x_1)+\mathcal{D}_p^{\alpha_m}M^P(x_1)\right]+f_t(x_1)=0
\end{equation}
subject to the following boundary conditions at $x_1\in\{0,L\}$:
\begin{subequations}
    \label{eq:bondary_condition_converse} 
    \begin{equation}
	\label{eq:axial_BC_converse_sub}
        N(x_1)=0~\text{or}~\delta{u_0}(x_1)=0
    \end{equation}
    \begin{equation}
	\label{Transverse_BC1_converse_sub}
        M(x_1)=0~\text{or}~\delta\left[\frac{\mathrm{d}{w_0}(x_1)}{\mathrm{d}x_1}\right]=0
    \end{equation}
    \begin{equation}
	\label{Transverse_BC2_converse_sub}
        \frac{\mathrm{d}M(x_1)}{\mathrm{d}x_1}=0~\text{or}~\delta{w_0}(x_1)=0
    \end{equation}
and following conditions corresponding to piezoelectric patch at $x_1\in\{x_0,x_0+L_P\}$: 
    \begin{equation}
	\label{eq:axial_BC_converse_piezo}
        N^P(x_1)=0~\text{or}~\delta{u_0}(x_1)=0
    \end{equation}
    \begin{equation}
	\label{Transverse_BC1_converse_piezo}
        M^P(x_1)=0~\text{or}~\delta\left[\frac{\mathrm{d}{w_0}(x_1)}{\mathrm{d}x_1}\right]=0
    \end{equation}
    \begin{equation}
	\label{Transverse_BC2_converse_piezo}
        \frac{\mathrm{d}M^P(x_1)}{\mathrm{d}x_1}=0~\text{or}~\delta{w_0}(x_1)=0
    \end{equation}
\end{subequations}
In the above equations,  ${_{x_1-l_B}^{R-RL}}D_{x_1+{l_A}}^{\alpha}(.)$ is a Riesz-type Riemann–Liouville fractional-order derivative defined, analogous to Eq.~\eqref{eq:RC_def}, as:
\begin{equation}
    \label{eq:R-RL_derivative}
    \mathcal{D}^{\alpha}(\cdot) = \frac{1}{2}\Gamma(2-\alpha)\left[l_B^{\alpha-1}\left(_{x_1-l_B}^{RL}D_{x_1}^{\alpha}(.)\right) - l_A^{\alpha-1}\left(_{x_1}^{RL}D_{x_1+l_A}^{\alpha}(.)\right)\right]
\end{equation}
where $_{x_1-l_B}^{RL}D_{x_1}^{\alpha}(.)$ is the left-Riemann Liouville fractional derivative of order $\alpha$ and $_{x_1}^{RL}D_{x_1+l_A}^{\alpha}(.)$ is the right- Riemann–Liouville fractional derivative of identical order \cite{podlubny1998fractional}. Subscript $p$ is used in the above for derivatives corresponding to the piezoelectric patch.
	
Here, the axial and bending stress resultants within the substrate are defined as 
\begin{subequations}
    \label{eq:stress_resultant_def}
    \begin{equation}
	\label{eq:stress_resultant_sub}
        N(x_1)={\int_0^b}{{\int_{-h/2}^{h/2}}\Tilde{\sigma}_{11}\,\mathrm{d}x_3}\,\mathrm{d}x_2~~~~~~~~~~~M(x_1)={\int_0^b}{{\int_{-h/2}^{h/2}}{x_3}\Tilde{\sigma}_{11}\,\mathrm{d}x_3}\,\mathrm{d}x_2  
    \end{equation}
where $\tilde{\sigma}_{11}$ is the stress within the nonlocal substrate given in Eq.~\eqref{eq:subs_const_relations}. Similarly, the stress resultants in the piezoelectric patch are defined as follows:
    \begin{equation}
	\label{eq:stress_resultant_piezo}
        N^P(x_1)=\begin{cases}
            {\int_0^b}{{\int_{h/2}^{h/2+h_P}}\Tilde{\sigma}^P_{11}\,\mathrm{d}x_3}\,\mathrm{d}x_2 & \text{for~}x_1\in(x_0,x_0+L_P)\\
            0 & \text{elsewhere}
        \end{cases}
        \end{equation}
        \begin{equation}
	\label{eq:stress_resultant_piezo2}
        M^P(x_1)=\begin{cases}
            {\int_0^b}{{\int_{h/2}^{h/2+h_P}}{x_3}\Tilde{\sigma}^P_{11}\,\mathrm{d}x_3}\,\mathrm{d}x_2  & \text{for~}x_1\in(x_0,x_0+L_P)\\
            0 & \text{elsewhere}
        \end{cases}
        \end{equation}
    here $\tilde{\sigma}^P_{11}$ is the stress within the nonlocal piezoelectric layer as given in Eq.~\eqref{eq:subs_const_relations}. Note that the mechanical force due to electrical loading on the nonlocal piezoelectric layer is captured in the stress resultants $N^P(x_1)$ and $M^{P}(x_1)$. 
\end{subequations}

\subsection{Governing differential equations for direct piezoelectric effect}	
In this section, we develop the governing equations of equilibrium for the direct piezoelectric effect within a nonlocal smart beam. More clearly, the nonlocal piezoelectric layer presents an electrical response to mechanical loading. Therefore, in this case, both the mechanical displacement and electric potential are independent field variables and require coupled electro-mechanical governing equations to be developed for all these field variables. 
	
The electro-mechanical governing equations of equilibrium for the smart beam demonstrating direct piezoelectric coupling are given as follows:
\begin{subequations}
    \label{eq:gov_diff_eq_direct}
    \begin{equation}
	\label{eq:axial_GDE_direct} 
        \mathcal{D}^{\alpha_m}N(x_1)\mathcal{D}^{\alpha_m}_{p}N^P(x_1)+f_a(x_1)=0~~~\frac{\mathrm{d}}{\mathrm{d}x_1}\left[\mathcal{D}^{\alpha_m}M(x_1)+\mathcal{D}_p^{\alpha_m}M^P(x_1)\right]+f_t(x_1)=0
    \end{equation}
    \begin{equation}
	\label{eq:electric_GDE}
	-\frac{\tilde{P}_{3}(x_1)}{h_P}=0
    \end{equation}
\end{subequations}
and subject to the boundary conditions derived previously in \eqref{eq:bondary_condition_converse}.
Additionally, the electrical field resultant is defined as:
\begin{equation}
    \label{eq:elec_field_stress_resultant}
    \Tilde{P}_{3}(x_1)={\int_0^b}{{\int_{h/2}^{h/2+h_P}}\Tilde{D}_{3}\,\mathrm{d}x_3}\,\mathrm{d}x_2,~~~~\forall~~x_1\in(x_0,x_0+L_P)
\end{equation}
where the electrical displacement field in the nonlocal piezoelectric layer is as defined in Eq.~\eqref{eq:piezo_mat_cons}. 
	
	
\section{Finite Element Model} 
\label{sec:FEM}
In the above sections, we derive the constitutive relations and governing differential equations for the nonlocal smart beam. Due to the integro-differential nature of the governing equations presented in Eqs.~\eqref{eq:gov_diff_eq_converse} and \eqref{eq:gov_diff_eq_direct}, along with the presence of multiphysics coupling in these equations, we develop a numerical model to solve them. To achieve this, we expand upon the fractional-Finite Element Method initially developed for nonlocal elasticity in \cite{patnaik2020ritz}, extending it to a multiphysics framework.

Before we start with developing the numerical model of the nonlocal smart beam for converse and direct piezoelectric coupling, we revisit the potential energy expression derived earlier. We recast the potential energy given in Eq.~\eqref{eq:poten_eng} using kinematic (strain-displacement) and material (stress-strain) constitutive relations in Eqs.~\eqref{eq:elec_fields}-\eqref{eq:subs_const_relations} as:
\begin{equation}
    \begin{aligned}
    \label{eq:weak_form}
    \bm{\Pi} = &\frac{1}{2}{\int_0^{L}E_SA\left(D_{x_1}^{\alpha_m}{u_0}\right)^2\,\mathrm{d}{x_1}} + \frac{1}{2}{\int_0^{L}E_SI\left(D_{x_1}^{\alpha_m}\left[\frac{\mathrm{d}{w_0}}{\mathrm{d}{{x_1}}}\right]\right)^2\,\mathrm{d}{x_1}} + \frac{1}{2}{\int_{x_0}^{x_0+L_P}E_PA_P\left(D_{x_1}^{\alpha_m}{u_0}\right)^2\,\mathrm{d}{x_1}}\\ &+ \frac{1}{2}{\int_{x_0}^{x_0+L_P}E_PI_P\left(D_{x_1}^{\alpha_m}\left[\frac{\mathrm{d}{w_0}}{\mathrm{d}{{x_1}}}\right]\right)^2\,\mathrm{d}{x_1}} - {\int_{x_0}^{x_0+L_P}E_PB_P\left(D_{x_1}^{\alpha_m}{u_0}\right)\left(D_{x_1}^{\alpha_m}\left[\frac{\mathrm{d}{w_0}}{\mathrm{d}{{x_1}}}\right]\right)\,\mathrm{d}{x_1}}  \\& -\int_{x_0}^{x_0+L_P}\frac{\varepsilon_{31}{\phi_0}{A_P}}{h_P}\left(D_{x_1}^{\alpha_m}{u_0}\right)\,\mathrm{d}{x_1}+ {\int_{x_0}^{x_0+L_P}\frac{\varepsilon_{31}{\phi_0}{B_P}}{h_P}\left(D_{x_1}^{\alpha_m}\left[\frac{\mathrm{d}{w_0}}{\mathrm{d}{{x_1}}}\right]\right)\,\mathrm{d}{x_1}} \\& -\frac{1}{2}\int_{x_0}^{x_0+L_P}\frac{a_{33}A_P}{{h_P}^2}\left({\phi_0}^2\right)\,\mathrm{d}{x_1}  -\int_{0}^{L}{{f_a}}{u_0}\,\mathrm{d}{x_1}\  - \int_{0}^{L}{{f_t}}{w_0}\,\mathrm{d}{x_1}\ 
    \end{aligned}
\end{equation}
where $A$ and $I$ are the cross-sectional area and moment of inertia of the substrate beam, respectively.  Similarly, for the piezoelectric patch, $A_P$, $B_P$, and $I_P$ are given as:
\begin{equation}
    \label{eq:ABD_matrix}
    A_P = bh_P,~~~B_P = \frac{b}{2}(h_P^2+hh_P),~~~I_P = \frac{b}{3}\left(\frac{3}{4}h^2h_P+\frac{3}{2}hh_P^2+ h_P^3\right)
\end{equation}
where $h$ is the thickness of the substrate beam, $h_P$ is the thickness of the piezoelectric patch, and $b$ is the width of the smart structure. Applying the first variation over the above expression would provide:
\begin{equation}
    \begin{aligned}
    \label{eq:First_variation_of_weak_form}
    &\delta\bm{\Pi} = {\int_0^{L}E_SA\left(D_{x_1}^{\alpha_m}{u_0}\right)\left(D_{x_1}^{\alpha_m}\delta{u_0}\right)\,\mathrm{d}{x_1}} + {\int_0^{L}E_SI\left(D_{x_1}^{\alpha_m}\left[\frac{\mathrm{d}{w_0}}{\mathrm{d}{{x_1}}}\right]\right)\left(D_{x_1}^{\alpha_m}\left[\frac{\mathrm{d}\delta{w_0}}{\mathrm{d}{{x_1}}}\right]\right)\,\mathrm{d}{x_1}} \\& +{\int_{x_0}^{x_0+L_P}E_PA_P\left(D_{x_1}^{\alpha_m}{u_0}\right)\left(D_{x_1}^{\alpha_m}\delta{u_0}\right)\,\mathrm{d}{x_1}}+{\int_{x_0}^{x_0+L_P}E_PI_P\left(D_{x_1}^{\alpha_m}\left[\frac{\mathrm{d}{w_0}}{\mathrm{d}{{x_1}}}\right]\right)\left(D_{x_1}^{\alpha_m}\left[\frac{\mathrm{d}\delta{w_0}}{\mathrm{d}{{x_1}}}\right]\right)\,\mathrm{d}{x_1}} \\& - {\int_{x_0}^{x_0+L_P}E_PB_P\left(D_{x_1}^{\alpha_m}{u_0}\right)\left(D_{x_1}^{\alpha_m}\left[\frac{\mathrm{d}\delta{w_0}}{\mathrm{d}{{x_1}}}\right]\right)\,\mathrm{d}{x_1}} - {\int_{x_0}^{x_0+L_P}E_PB_P\left(D_{x_1}^{\alpha_m}\left[\frac{\mathrm{d}{w_0}}{\mathrm{d}{{x_1}}}\right]\right)\left(D_{x_1}^{\alpha_m}\delta{u_0}\right)\,\mathrm{d}{x_1}} \\& -\int_{x_0}^{x_0+L_P}\frac{\varepsilon_{31}{\phi_0}{A_P}}{h_P}\left(D_{x_1}^{\alpha_m}\delta{u_0}\right)\,\mathrm{d}{x_1} - \int_{x_0}^{x_0+L_P}\frac{\varepsilon_{31}{A_P}}{h_P}\left(D_{x_1}^{\alpha_m}{u_0}\right)\left(\delta{\phi_0}\right)\,\mathrm{d}{x_1} \\& + {\int_{x_0}^{x_0+L_P}\frac{\varepsilon_{31}{\phi_0}{B_P}}{h_P}\left(D_{x_1}^{\alpha_m}\left[\frac{\mathrm{d}\delta{w_0}}{\mathrm{d}{{x_1}}}\right]\right)\,\mathrm{d}{x_1}} +{\int_{x_0}^{x_0+L_P}\frac{\varepsilon_{31}{B_P}}{h_P}\left(D_{x_1}^{\alpha_m}\left[\frac{\mathrm{d}{w_0}}{\mathrm{d}{{x_1}}}\right]\right)\left(\delta{\phi_0}\right)\,\mathrm{d}{x_1}} \\& -\int_{x_0}^{x_0+L_P}\frac{a_{33}A_P}{{h_P}^2}\left({\phi_0}\right)\left(\delta{\phi_0}\right)\,\mathrm{d}{x_1} -\int_{0}^{L}{{f_a}}\delta{u_0}\,\mathrm{d}{x_1}\ - \int_{0}^{L}{{f_t}}\delta{w_0}\,\mathrm{d}{x_1}\ 
    \end{aligned}
\end{equation}
where $\delta u_0$ and $\delta w_0$ are variations in the axial and transverse displacements, respectively, and $\delta \phi_0$ is the variation in electric potential. Recall that for the nonlocal smart beam demonstrating converse piezoelectric coupling, the electric potential is a prescribed variable ($\delta \phi_0=0$). 

To develop the finite element model, the entire length of the smart beam, including the piezoelectric patch, is divided into $\mathcal{N}-$ uniform, one-dimensional two-noded elements. These elements ${N}_e\in{1,2...\mathcal{N}}$ are defined such that $\cup_{i=1}^{N_e}=\mathcal{N}$ and $N_{j}\cap N_{k}=\emptyset,~~j\neq k$. The equivalent single-layer model for the smart beam ensures a single 1D element (through the thickness) for the substrate and the piezoelectric patch. 
The equivalent single-layer model for the smart beam ensures a single 1D element (through the thickness) for the substrate and the piezoelectric patch. Following Eq.~\eqref{eq:First_variation_of_weak_form}, we choose a $C^0-$continuous (Lagrange) shape function to interpolate the axial displacement and electrical potential and a $C^1-$continuous (Hermite) shape function for the transverse displacement. 

The field variable at any point in the domain is now given as:
\begin{equation}
    \label{eq:matrix_form_Disp_u}
    {u_0}({x_1})=[\mathbf{L}(x_1)]\{\mathbf{u_e}(x_1)\},~~~{\phi_0}({x_1})=[\mathbf{L}(x_1)]\{\mathbf{\phi_e}(x_1)\},~~~{w_0}({x_1})=[\mathbf{H}(x_1)]\{\mathbf{w_e}(x_1)\}
\end{equation}
where $[\mathbf{L}(x_1)]$ and $[\mathbf{H}(x_1)]$ are the Lagrange and Hermite shape function matrices, respectively, and $\{\mathbf{u_e}(x_1)\}$, $\{\mathbf{w_e}(x_1)\}$ and $\{\mathbf{\phi_e}(x_1)\}$ are element vectors for corresponding nodal values. 
Note that these nodal vectors correspond to the element containing the point $x_1$. 
The integer-order derivatives of the displacement can be written as:
\begin{subequations}
\label{eq: integer_der_intpl_full}
 \begin{equation}
    \label{eq: integer_der_intpl}
     \frac{\mathrm{d}u_0(x_1)}{\mathrm{d}x_1}=[B_u]\{\mathbf{u_e}(x_1)\},~~~\frac{\mathrm{d}^2w_0(x_1)}{\mathrm{d}x_1^2}=[B_w]\{\mathbf{w_e}(x_1)\}
\end{equation}
where
\begin{equation}
    \label{eq: integer_der2_intpl}
     [B_u]=\frac{\mathrm{d}}{\mathrm{d}x_1}[\mathbf{L}(x_1)]~~~[B_w]=\frac{\mathrm{d}^2}{\mathrm{d}x_1^2}[\mathbf{H}(x_1)]
\end{equation}   
\end{subequations}
It is clear from the above expression that the integer-order derivatives at any point $x_1$ can be expressed only in terms of the nodal values of the corresponding element. However, this is not the case for fractional-order derivatives that require the information across the horizon of nonlocal influence to be included. Following the interpolation provided in Eq.~\eqref{eq:matrix_form_Disp_u}, the fractional-order derivatives in Eq.~\eqref{eq:First_variation_of_weak_form} can be written as:
\begin{equation}
    \label{eq:dervative_Disp_u}
    D_{x_1}^{\alpha_m}{{u_0}({x_1})}=[\mathbf{B}{_u^{\alpha_m}}(x_1)]\{\mathbf{u}_g\},~~~D_{x_1}^{\alpha_m}\left[\frac{\mathrm{d}{w_0(x_1)}}{\mathrm{d}{{x_1}}}\right]=[\mathbf{B}{_w^{\alpha_m}}(x_1)]\{\mathbf{w}_g\}
\end{equation}
where $\{\mathbf{u}_g\}$ and $\{\mathbf{w}_g\}$ are the global vectors for nodal displacements obtained via appropriate assembly of the corresponding element vectors. Matrixes $[\mathbf{B}{_u^{\alpha_m}}(x_1)]$ and $[\mathbf{B}{_w^{\alpha_m}}(x_1)]$ are strain-displacement matrices for nonlocal interaction defined as fractional-order derivatives of interpolation functions. These matrices are given as\cite{patnaik2020ritz}:
\begin{subequations}
    \begin{equation}
	\label{eq:nonlocal_axial_disp}
        D_{x_1}^{\alpha_m}{{u_0}({x_1})}=[\mathbf{B}{_u^{\alpha_m}}(x_1)]\{\mathbf{u}_g\}
    \end{equation}
    where,
    \begin{equation}
	\label{eq:B_alpha_mat}
        [\mathbf{B}{_u^{\alpha_m}}(x_1)]=\int_{x_1-l_A}^{x_1+l_B}A(x_1, s_1, \alpha_m, l_A, l_B)[B_u][\mathcal{C}(s_1)]\,\mathrm{d}{s_1}
    \end{equation}
    where $[\mathcal{C}(s_1)]$ is a boolean matrix to ensure appropriate assembly of element matrices for $s_1$ within the global system matrices.
    \end{subequations}
A detailed derivation of these expressions is available in literature\cite{patnaik2020ritz,sidhardh2020geometrically}, and not repeated here for the sake of brevity. Interested readers may refer to a brief description of the steps involved in this derivation provided in the Appendix.

\subsection{Converse Piezoelectric Effect}
\label{sec:Converse Piezoelectric Effect}
We start by developing an FE model for the nonlocal smart beam where the piezoelectric patch is employed as an actuator. More clearly, the nonlocal smart beam is mechanically actuated by an electrical potential via a converse piezoelectric effect. Therefore, the electrical potential over the piezoelectric layer is prescribed ($\delta \phi_0=0$). 

We begin by employing the numerical approximations for fractional-order derivatives developed above in Eq.~\eqref{eq:First_variation_of_weak_form}. Subsequently, applying the principle of minimum potential energy ($\delta \Pi=0$), the algebraic governing equations of equilibrium can be written as follows: 
\begin{subequations}
    \label{eq:algebraic_gov_eq}
    \begin{equation}
	\label{eq:algebraic_gov_eq_u}
        [\mathbf{K_{uu}}]\{\mathbf{u}_g\}+[\mathbf{K_{uw}}]\{\mathbf{w}_g\} = \{\mathbf{F_{ae}}\}+\{\mathbf{F_{a}}\}
    \end{equation}
    \begin{equation}
	\label{eq:dervative_Disp_w}
        [\mathbf{K_{uw}}]^T\{\mathbf{u}_g\}+[\mathbf{K_{ww}}]\{\mathbf{w}_g\} = \{\mathbf{F_{te}}\}+\{\mathbf{F_{t}}\}
    \end{equation}
\end{subequations}
here $[\mathbf{K_{uu}}]$, $[\mathbf{K_{ww}}]$ and $[\mathbf{K_{uw}}]$  are the consolidated global stiffness matrices for the substrate beam and the piezoelectric patch. $\{\mathbf{F_{a}}\}$ and $\{\mathbf{F_{t}}\}$ are the mechanical forces due to the mechanical loads on the beam. $\{\mathbf{F_{ae}}\}$ and $\{\mathbf{F_{te}}\}$ are the mechanical force vectors due to the electrical potential applied over the beam via the converse piezoelectric effect. The expressions for all the consolidated stiffness matrices and force vectors are given as follows:
\begin{subequations}
    \label{eq:element_stiffness_mat}
    \begin{equation}
        [\mathbf{K_{uu}}]=[\mathbf{K_{uu}^s}]+[\mathbf{K_{uu}^p}]
    \end{equation}
    \begin{equation}
        [\mathbf{K_{ww}}]=[\mathbf{K_{ww}^s}]+[\mathbf{K_{ww}^p}]
    \end{equation}
    \begin{equation}
	\label{eq:element_stiffness_mat_u_subs}
        [\mathbf{K_{uu}^s}] = \int_{0}^{L}[\mathbf{B}_u^{\alpha_m}(x_1)]^T(E_SA)[\mathbf{B}_u^{\alpha_m}(x_1)]\,\mathrm{d}{x_1}
    \end{equation}
    \begin{equation}
	\label{eq:element_stiffness_mat_u_piezo}
        [\mathbf{K_{uu}^p}] = \int_{x_0}^{x_0+L_P}\left[(\mathbf{B}_u^{\alpha_m}(x_1))^P\right]^T(E_PA_P)\left[(\mathbf{B}_u^{\alpha_m}(x_1))^P\right]\,\mathrm{d}{x_1}
    \end{equation}
    \begin{equation}
	\label{eq:element_stiffness_mat_w_subs}
        [\mathbf{K_{ww}^s}] = \int_{0}^{L}[\mathbf{B}_w^{\alpha_m}(x_1)]^T(E_SI)[\mathbf{B}_w^{\alpha_m}(x_1)]\,\mathrm{d}{x_1}
    \end{equation}
   \begin{equation}
	\label{eq:element_stiffness_mat_w_piezo}
        [\mathbf{K_{ww}^p}] = \int_{x_0}^{x_0+L_P}\left[(\mathbf{B}_w^{\alpha_m}(x_1))^P\right]^T(E_PI_P)\left[(\mathbf{B}_w^{\alpha_m}(x_1))^P\right]\,\mathrm{d}{x_1}
    \end{equation}
    \begin{equation}
	\label{eq:element_stiffness_mat_u-w_piezo}
        [\mathbf{K_{uw}}] = \int_{x_0}^{x_0+L_P}\left[(\mathbf{B}_u^{\alpha_m}(x_1))^P\right]^T(-E_PB_P)\left[(\mathbf{B}_w^{\alpha_m}(x_1))^P\right]\,\mathrm{d}{x_1}
    \end{equation}
 and the force vectors are given as:
    \begin{equation}
        \begin{aligned}
        \label{eq:elec_force_vector_u_w}  
            &&\{\mathbf{F_{ae}}\}=\int_{x_0}^{x_0+L_P}\left(-\frac{\varepsilon_{31}\phi_0 A_P}{h_P}\right)\left[(\mathbf{B}_u^{\alpha_m}(x_1))^P\right]^T\,\mathrm{d}{x_1}
            \\&& \{\mathbf{F_{te}}\} = \int_{x_0}^{x_0+L_P}\left(\frac{\varepsilon_{31}\phi_0 B_P}{h_P}\right)\left[(\mathbf{B}_w^{\alpha_m}(x_1))^P\right]^T\,\mathrm{d}{x_1}  
        \end{aligned}        
    \end{equation}
    \begin{equation}
	\label{eq:mech_force_vector_u}  
	\{\mathbf{F_{a}}\}=\int_{0}^{L}f_{a}(x_1)[\mathbf{L}{(x_1)}]^T\,\mathrm{d}{x_1},~~
	\{\mathbf{F_{t}}\} = \int_{0}^{L}f_{t}(x_1)[\mathbf{H}{(x_1)}]^T\,\mathrm{d}{x_1} 
    \end{equation}
\end{subequations}
Recall from our previous discussion that the length scales for nonlocal influence at a point $x_1$ in the piezoelectric patch may differ from that of the same point in the substrate. Moreover, as seen in Eq.~\eqref{eq:B_alpha_mat} the nonlocal $[\mathbf{B}^{\alpha}]$ matrix depends on these length scales. Therefore, we use the superscript $[\cdot]^P$ in $\left[(\mathbf{B}_u^{\alpha_m}(x_1))^P\right]$ and $\left[(\mathbf{B}_w^{\alpha_m}(x_1))^P\right]$ to denote the fractional-order strain displacement matrices of piezoelectric solid.
	
\subsection{Direct Piezoelectric Effect}
Here, we develop the FE model for the nonlocal smart beam where the piezoelectric patch demonstrates a direct piezoelectric effect. More clearly, the piezoelectric patch presents electrical output when subjected to a mechanical load. Therefore, the three mid-plane variables $u_0$, $w_0$ and $\phi_0$ in Eq.~\eqref{eq:First_variation_of_weak_form} are all independent. Using the numerical approximations for fractional-order derivatives developed above, the weak form can be recast in terms of nodal degrees of freedom $\{\mathbf{u}_g\}, $$\{\mathbf{w}_g\}$ and $\{\bm{\phi}_g\}$. Finally, applying the principle of minimum potential energy ($\delta \Pi=0$), the algebraic governing equations of equilibrium can be written as follows: 	
\begin{subequations}
    \label{eq:algebraic_gov_eq_sensing}
    \begin{equation}
	\label{eq:algebraic_gov_eq_u_sensing}
        [\mathbf{K_{uu}}]\{\mathbf{u}_g\} +[\mathbf{K_{uw}}]\{\mathbf{w}_g\} + [\mathbf{K_{u\bm{\phi}}}]\{\bm{\phi}_g\} = \{\mathbf{F_{a}}\}
    \end{equation}
    \begin{equation}
	\label{eq:algebraic_gov_eq_w_sensing}
        [\mathbf{K_{uw}}]^T\{\mathbf{u}_g\}+[\mathbf{K_{ww}}]\{\mathbf{w}_g\} +[\mathbf{K_{w\bm{\phi}}}]\{\bm{\phi}_g\} = \{\mathbf{F_{t}}\}
    \end{equation}
    \begin{equation}
	\label{eq:algebraic_gov_eq_phi_sensing}
        [\mathbf{K_{u\bm{\phi}}}]^T\{\mathbf{u}_g\}+[\mathbf{K_{w\bm{\phi}}}]^T\{\mathbf{w}_g\}+[\mathbf{K_{\bm{\phi\phi}}}]\{\bm{\phi}_g\}=\{\textbf{0}\}
    \end{equation}
\end{subequations}
here $[\mathbf{K_{uu}}]$, $[\mathbf{K_{uw}}]$, $[\mathbf{K_{u\bm{\phi}}}]$, $[\mathbf{K_{ww}}]$, $[\mathbf{K_{w{\bm{\phi}}}}]$, and $[\mathbf{K_{\bm{\phi\phi}}}]$ are the consolidated global stiffness matrices for the substrate beam and the piezoelectric patch. $\{\mathbf{F_{a}}\}$ and $\{\mathbf{F_{t}}\}$ are the axial, transverse, and electrical potential force vectors. The expression for $[\mathbf{K_{uu}}]$, $[\mathbf{K_{uw}}]$ and $[\mathbf{K_{ww}}]$ are provided in Eq.~\eqref{eq:element_stiffness_mat}. The remaining terms are given as follows:
\begin{subequations}
    \label{eq:element_stiffness_mat_sensing}
    \begin{equation}
        \begin{aligned}
        \label{eq:element_stiffness_mat_u-w_sensing}
            &&[\mathbf{K_{u\bm{\phi}}}]=\int_{x_0}^{x_0+L_P}\left[(\mathbf{B}_u^{\alpha_m}(x_1))^P\right]^T\left(\frac{{A_P}{\varepsilon_{31}}}{h_P}\right)[\mathbf{L}(x_1)]\,\mathrm{d}{x_1}\\&&
            [\mathbf{K_{w{\bm{\phi}}}}]=\int_{x_0}^{x_0+L_P}\left[(\mathbf{B}_w^{\alpha_m}(x_1))^P\right]^T\left(\frac{-{B_P} {\varepsilon_{31}}}{h_P}\right)[\mathbf{L}(x_1)]\,\mathrm{d}{x_1}  
        \end{aligned}
    \end{equation}
    \begin{equation}
	\label{eq:element_stiffness_mat_w-phi_sensing}
        [\mathbf{K_{\bm{\phi\phi}}}]=\int_{x_0}^{x_0+L_P}[\mathbf{L}(x_1)]^T\left(-\frac{{A_P} {a_{33}}}{h_P^2}\right)[\mathbf{L}(x_1)]\,\mathrm{d}{x_1}
    \end{equation}
\end{subequations}

\section{Results and discussion}
\label{sec: results}    
We undertake numerical investigations into the multiphysics response of the nonlocal smart beam as predicted by the fractional-order model proposed here. In the numerical investigations conducted here, the length of the substrate beam is $L = 24.53~\mathrm{mm}$. We consider two cases: (i) a piezoelectric layer over the entire length of the substrate, $(x_0=0, L_P=L;~\text{see Eq.~\eqref{eq:energy_piez}})$ and (ii) a piezoelectric patch of length $L_P=0.3L$ over left end of the smart beam ($x_0=0$). The piezoelectric layer/patch spans the entire width of the substrate beam, $b = 6.4~\mathrm{mm}$. The thickness of the substrate beam is $h=0.14~\mathrm{mm}$, and the piezoelectric patch is $h_P=0.05~\mathrm{mm}$. Note that these dimensions are in keeping with the slender beam assumptions of the Euler-Bernoulli beam displacement theory. It is assumed that the piezoelectric patch is made of PZT-5H with elastic modulus $E_P=60.6~\mathrm{GPa}$ while the substrate beam is assumed to be constructed of brass with elastic modulus $E_S=105~\mathrm{GPa}$. The piezoelectric constant $e_{31}=16.604~\mathrm{C/m^2}$ and piezoelectric permittivity $a_{33}=0.26~\times~10^{-7}~\mathrm{F/m}$ for PZT-5H. The material constants of the substrate and the piezoelectric patch provided above are the following  \cite{wang2013analysis}. 

In the subsequent analysis, we require the fractional-order constitutive parameters for the substrate and the piezoelectric layer/patch. More clearly, we require $\alpha_m$ and $h_l$ for nonlocal elasticity within the substrate and for nonlocal piezoelectricity within the piezoelectric patch. 
While the models developed here are general and can consider independent numerical values of fractional-order constitutive parameters for the substrate and the patch, we assume (for convenience) these parameters for the substrate and patch to be identical. Therefore, $\alpha_m$ for substrate and piezoelectric patch are identical ($=\alpha$).
Similarly, nonlocal length scales, for points sufficiently within the substrate/piezoelectric solid domain, are considered equal $l_A=l_B=h_l$\cite{patnaik2020ritz}. These constitutive parameters, fractional-order $\alpha$ and length scale $h_l$, for the nonlocal smart beam, will be provided where necessary. 
    
\begin{figure}[H]
    \centering
    \includegraphics[width=0.5\linewidth]{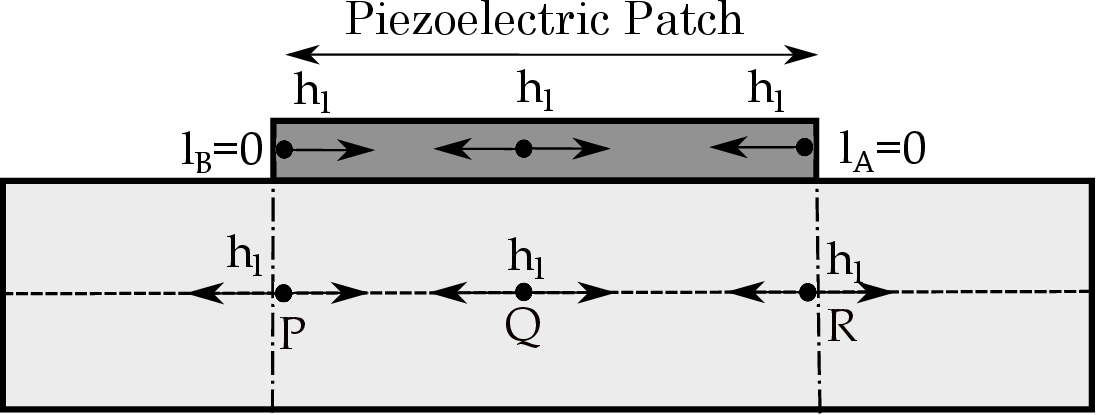}
    \caption{A schematic illustration of the nonlocal length scales within the piezoelectric patch and the substrate for the smart beam. Note that for the same point along the length of the smart beam, say P or R, the length scales in the piezoelectric patch and substrate are different. This renders differential nonlocal interactions across the piezoelectric patch and the substrate.}
    \label{fig:nonlocal_patch_intrection}
\end{figure}
Recall that the nonlocal length scales are appropriately truncated ($l_A\neq l_B$) for points close to external boundaries\cite{patnaik2020ritz}, either within the substrate or the piezoelectric solid. More clearly, for points closer to the boundaries of the substrate at $x_1=0,L$, and to the boundaries of the piezoelectric patch at $x_1=x_0,x_0+L_P$, the nonlocal length scales are defined such that $l_A\neq l_B$.  An illustration of the same is provided in the schematic given in Fig.~\ref{fig:nonlocal_patch_intrection}. In the subsequent, we quantitatively examine the influence of the parameters of fractional-order constitutive theory: $\alpha$ and $h_l$, on the response of the smart beam under various electrical and mechanical boundary conditions. 

 \subsection{Convergence}   
First, we establish the convergence of the f-FE model developed for the smart beam. Note that the factor $N_{inf}~(=h_l/l_e)$, where $l_e$ is the length of the discretized element, defined as the "dynamic rate of convergence" governs the convergence of successive integrations in the integro-differential expressions for fractional-order derivatives [35]. As a result, the numerical evaluation of fractional-order derivatives is compared for various numbers of elements within the nonlocal horizon of influence. 

\begin{figure}[H]
    \centering
    \begin{subfigure}{.5\textwidth}
        \centering
        \includegraphics[width=1\linewidth]{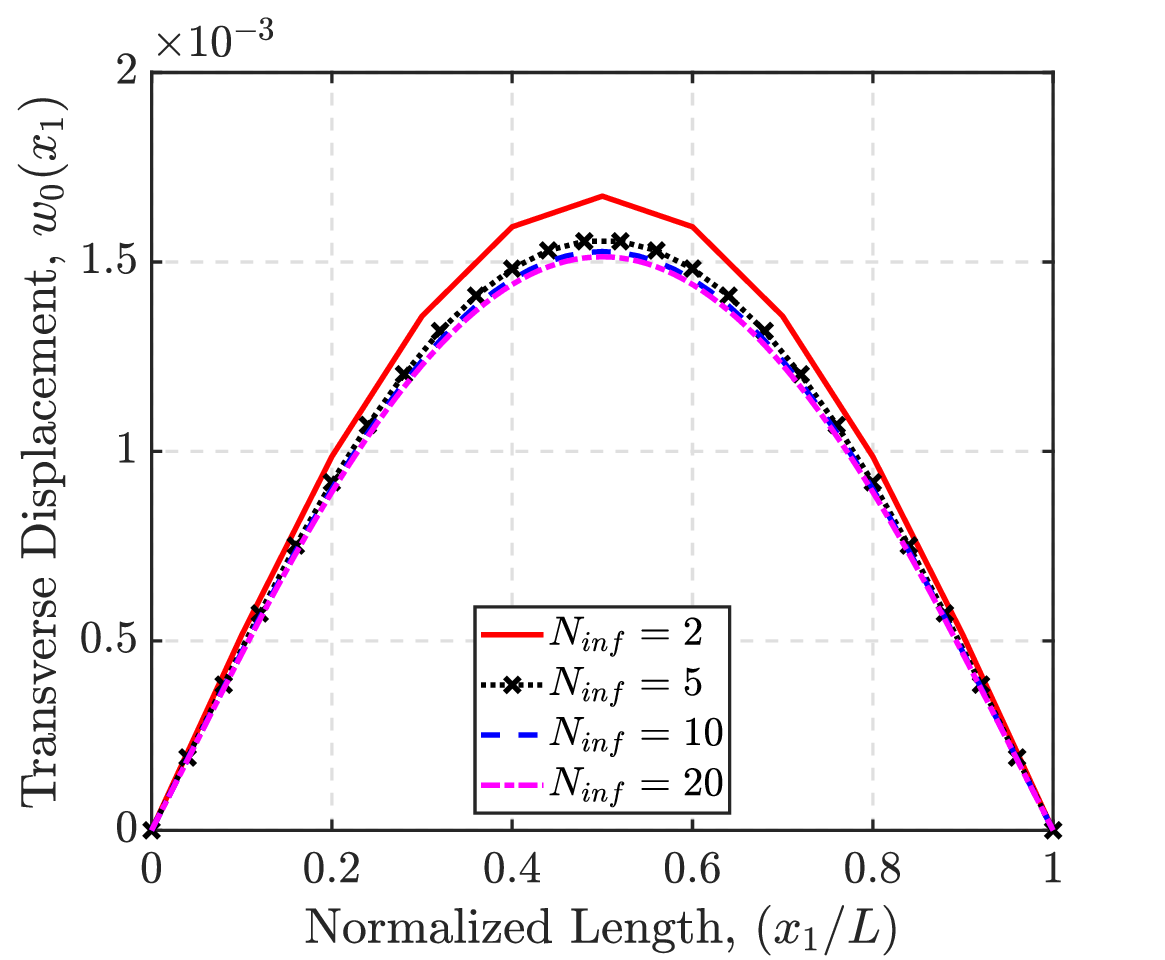}
        \caption{}
        \label{fig:convergence}
    \end{subfigure}%
    \begin{subfigure}{.5\textwidth}
	\centering
	\includegraphics[width=1\linewidth]{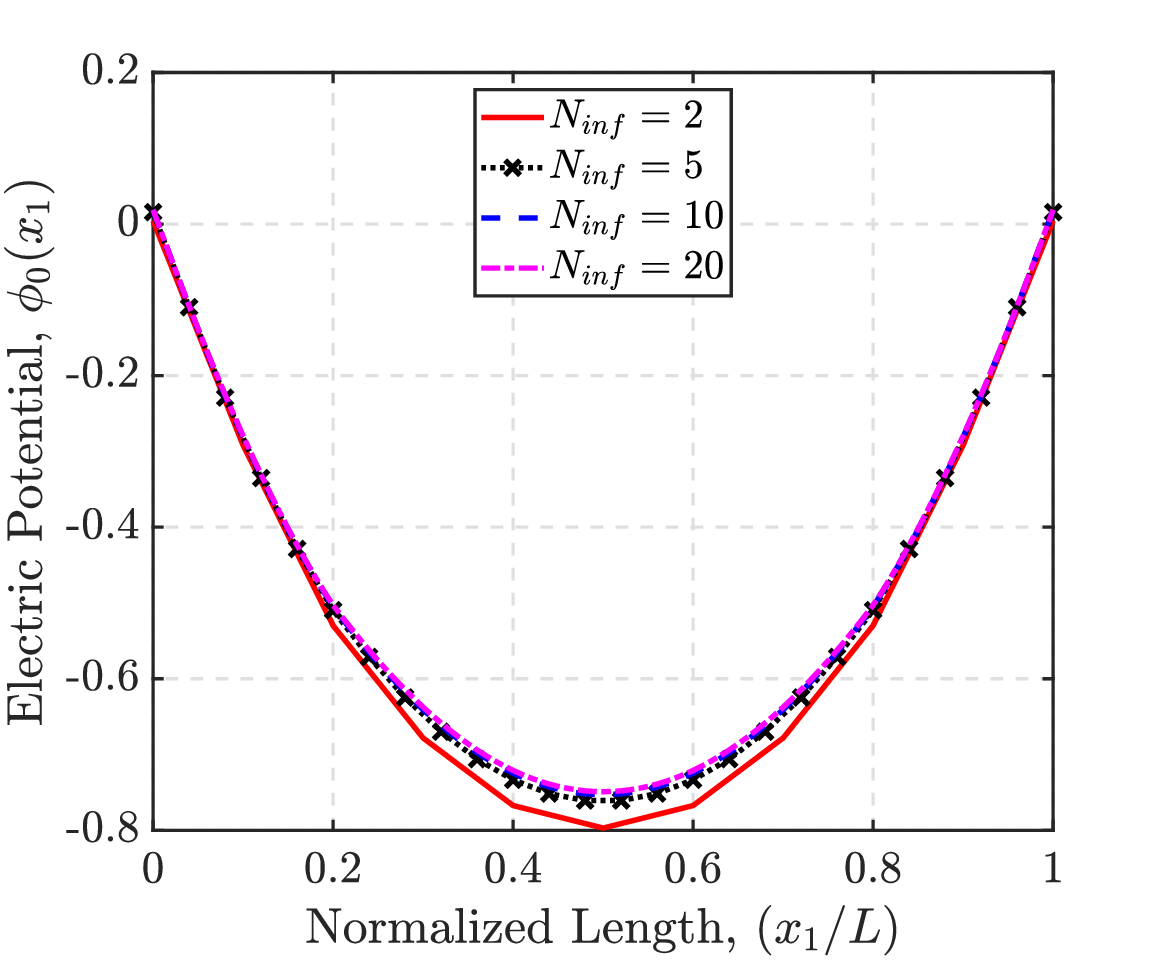}
	\caption{}
	\label{fig:convergence_elec}
    \end{subfigure}
    \caption{f-FEM solution for (a) converse, and (b) direct piezoelectric effect, in a nonlocal smart beam for the different number of elements. (Transverse displacement in ($\mathrm{m})$ and electric potential in ($\mathrm{V}$))}
    \label{fig:numerical convergence}
\end{figure}

For this study, we consider the example of a simply supported fractional-order smart beam, $\alpha=0.8$ and $h_l=L/5$, with a piezoelectric layer throughout the length of the beam. The smart beam is actuated by a mechanical load of $q_0(x_1)=100~\mathrm{N/m}$ and an electrical load $\phi=100\mathrm{V}$. The transverse displacement of the nonlocal smart beam evaluated with f-FEM for different values of $N_{inf}$ is compared in Fig.~\ref{fig:convergence}. Clearly, the f-FEM has converged ($<1\%$ difference) for a choice of $N_{inf}=10$. 


The above convergence pertains to a simply supported fractional-order smart beam, demonstrating a converse piezoelectric effect. Additionally, for the sake of completion, the convergence of the numerical model for fractional-order smart beam demonstrating direct piezoelectric effect is also provided in Fig.~\ref{fig:convergence_elec}. The nonlocal smart beam is subject to a mechanical load $q_0(x_1)=1~\mathrm{N/m}$ in this study. Clearly, the electrical potential generated within the piezoelectric layer, as predicted by the f-FE model has converged for $N_{inf}= 10$. This establishes an appropriate evaluation and the numerical convergence of the fractional-order derivatives for converse and direct piezoelectric effects. Therefore, this choice of the FE mesh will be considered in all the subsequent numerical studies unless mentioned otherwise. 

\subsection{Validation}	
We undertake a two-fold validation of the f-FE multiphysics coupled model developed here. First, the f-FE tool is validated for the numerical evaluation of fractional-order derivatives by comparison with a numerical analysis of nonlocal elasticity \cite{patnaik2020ritz}. Later, the multiphysics coupling and associated system matrices are validated with 3D FEA \cite{comsol}.\\

\noindent
\textbf{Validation \#1:} First, the f-FE tool developed above for a nonlocal smart beam is reduced for a nonlocal elastic beam. For this purpose, we consider the following assumptions over the f-FE model developed in Eq.~\eqref{eq:algebraic_gov_eq}: the potential difference across the thickness of the piezoelectric layer is zero ($\phi_0(x_1)=0$), the thickness of the piezoelectric layer is zero ($h_P=0$). The above assumptions reduce the system equations in Eq.~\eqref{eq:algebraic_gov_eq} to a nonlocal elastic beam studied in \cite{patnaik2020ritz}. The normalized maximum transverse displacement $(\Bar{w})$ of a clamped-clamped fractional-order Euler-Bernoulli beam subjected to a uniformly distributed load (UDL), $q_0$, is computed for different values of $\alpha$ and $h_l$, and the results are summarized in Table \ref{tab:Compression}. These numerical outcomes have been non-dimensionalized as follows: 
\begin{equation}
    \label{eq:normalizing factor}
   \Bar{w} = \frac{384EI}{q_0L^4}w_0\left(\frac{L}{2}\right)
\end{equation}
An excellent agreement between the results predicted by the f-FE model developed here with \cite{patnaik2020ritz} is demonstrated in Table~\ref{tab:Compression}. This attests to the efficacy of the f-FE solver developed here for an accurate numerical estimation of the fractional-order derivatives. 

\begin{table}[H]
\begin{center}
    \begin{tabular}{ c c c c c c } 
    \hline
    \hline
    $h_l$ & & \multicolumn{4}{c}{$\overline{w}$} \\ 
    \hline
    \hline
     & & $\alpha=1.0$ & $\alpha=0.9$ & $\alpha=0.8$ & $\alpha=0.7$ \\
     \hline
     \multirow{2}{*}{$L/10$} & Present & 1.0000 & 1.0243 & 1.0456 & 1.0673 \\[1.5ex]
     & \cite{patnaik2020ritz} & 1.0000 & 1.0243 & 1.0456 & 1.0673  \\[1.5ex]
     \hline
    \multirow{2}{*}{$L/5$} & Present & 1.0000 & 1.0720 & 1.1401 & 1.2098 \\[1.5ex]
     & \cite{patnaik2020ritz} & 1.0000 & 1.0720 & 1.1401 & 1.2098 \\    
    \hline
    \hline
    \end{tabular}
    \caption{Normalized transverse displacements for a clamped-clamped fractional-order Euler-Bernoulli elastic beam for constitutive parameters. Validation of the f-FEM via comparison with existing literature \cite{patnaik2020ritz}.}
    \label{tab:Compression}
\end{center}
\end{table}
\noindent
\textbf{Validation \#2:} We present here a comparison of the multifield static response predicted by the numerical code with commercial FEA. For this purpose, we consider $\alpha=1$ in the FE numerical code, to ignore nonlocal effects. We conduct a 3D FEA using commercial FEA package, COMSOL Multiphysics \cite{comsol}. The simulation employs COMSOL Multiphysics modules for solid mechanics and electrostatics. The 3D FEA considers identical geometric dimensions (length, width and height) and material properties for the substrate and piezoelectric material used in our numerical model ($\nu=0.2$). The substrate and piezoelectric materials are discretized using {tetrahedral} elements, and verified for convergence. Following this, a stationary study is conducted within COMSOL to determine the transverse deflection of the smart beam subject to cantilever boundary conditions, with PZT-5H layer placed on the top surface of the substrate beam.
\begin{figure}[H]
\color{blue}
    \centering
    \includegraphics[width=0.5\linewidth]{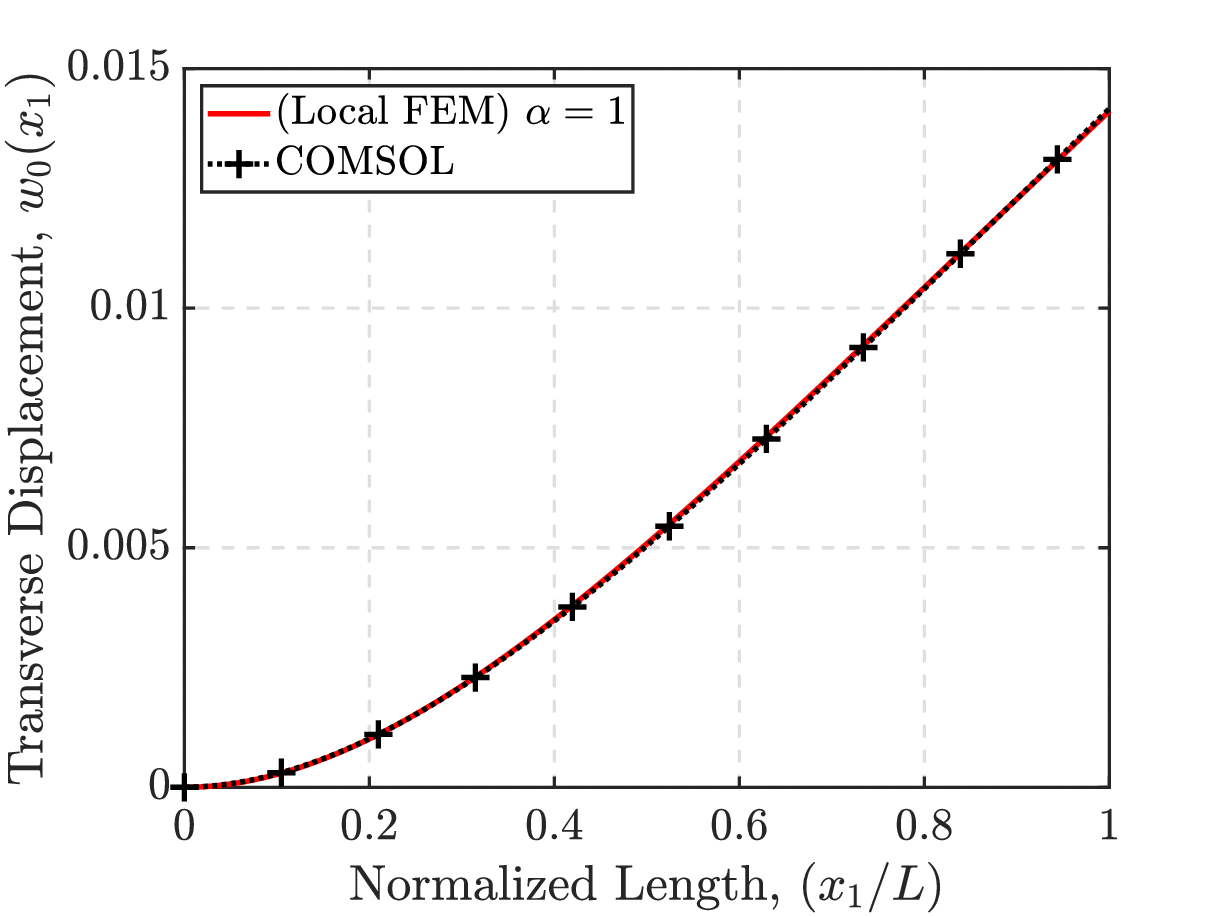}
    \caption{Transverse displacement (in $\mathrm{m}$) of the cantilever smart beam for $\alpha=1$ is compared with 3D FEA conducted in COMSOL multiphysics.}
    \label{fig:Validation_2_CF}
\end{figure}
The smart structure is also subject to a distributed mechanical load of 100 $\mathrm{N/m}$, and a uniform electric potential of $\phi=100~\mathrm{V}$ uniformly distributed across the length of the piezoelectric layer.  A comparison of the transverse displacement along the length of the beam as predicted by f-FEM (for $\alpha=1$) and 3D FEA is provided in Fig.~\ref{fig:Validation_2_CF}. An excellent agreement in the response, with less than $1\%$ difference of the transverse displacement at the free end, is noted between local elastic piezoelectric solid modeled via f-FEM ($\alpha=1$) and 3D FEA. This attests to an accurate estimation of the electro-mechanical coupling system matrices in Eq.~\ref{eq:algebraic_gov_eq} and Eq.~\ref{eq:algebraic_gov_eq_sensing} above.

\subsection{Parametric studies}
In this section, we conduct a series of parametric studies on nonlocal smart beams with varying fractional-order constitutive parameters. The intention of these studies is to investigate the effect of nonlocal (long-range) interactions within the structure over the elastic, electrical and coupled response of the smart beam. The effect of nonlocal interactions over the converse and direct piezoelectric coupling is intended to be realized by these parametric studies. Thereby, interesting insights into the role of nonlocal interactions towards tuning multiphysics coupling may be explored. The validated f-FE models for fractional-order smart Euler-Bernoulli beams, established above, will be used in all subsequent studies.
 
\subsubsection{Smart beam with piezoelectric layer}
We start with an analysis of the smart beam with a piezoelectric layer (throughout the length of the substrate). We investigate the effect of mechanical, electrical and combined loads on the multiphysics response of the smart beam. In this study, for the sake of brevity, we restrict ourselves to the simply-supported mechanical boundary conditions\cite{reddy2019introduction}. 

\paragraph{\textit{Converse piezoelectric effect}}~\\
\noindent
In this study, we consider the smart beam is subject to a uniformly distributed mechanical load $q_0(x_1)$, and/or a distributed electrical potential $\phi_0(x_1)$ at the top surface of the piezoelectric layer. The converse piezoelectric effect caused by the applied electrical potential will present a mechanical deformation of the smart beam.  The algebraic equations developed in Eq.~\eqref{eq:algebraic_gov_eq} are solved for these electro-mechanical loads to determine the response of the nonlocal smart beam.
We examine the effect of the fractional-order constitutive parameters ($\alpha$ and $h_l$) on the multiphysics response of the smart beam under various loading conditions. 

\begin{figure}[H]
    \centering
    \begin{subfigure}{.5\textwidth}
        \centering
        \includegraphics[width=1\linewidth]{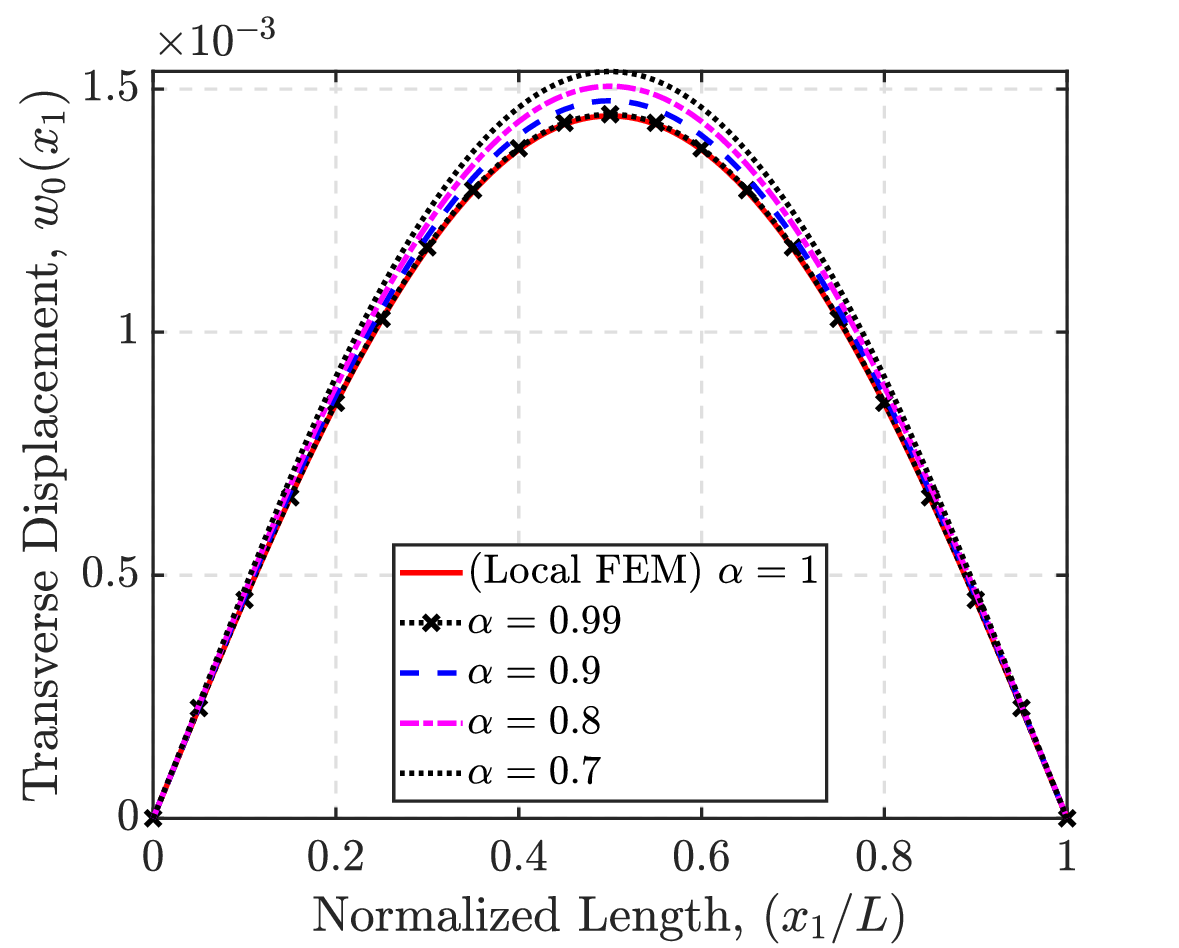}
        \caption{$w_0(x_1)$~vs~$\alpha$ for $h_l=L/5$}
        \label{fig:Result_2_SS_Actuation_WO_elec_load}
    \end{subfigure}%
    \begin{subfigure}{.5\textwidth}
	\centering
	\includegraphics[width=1\linewidth]{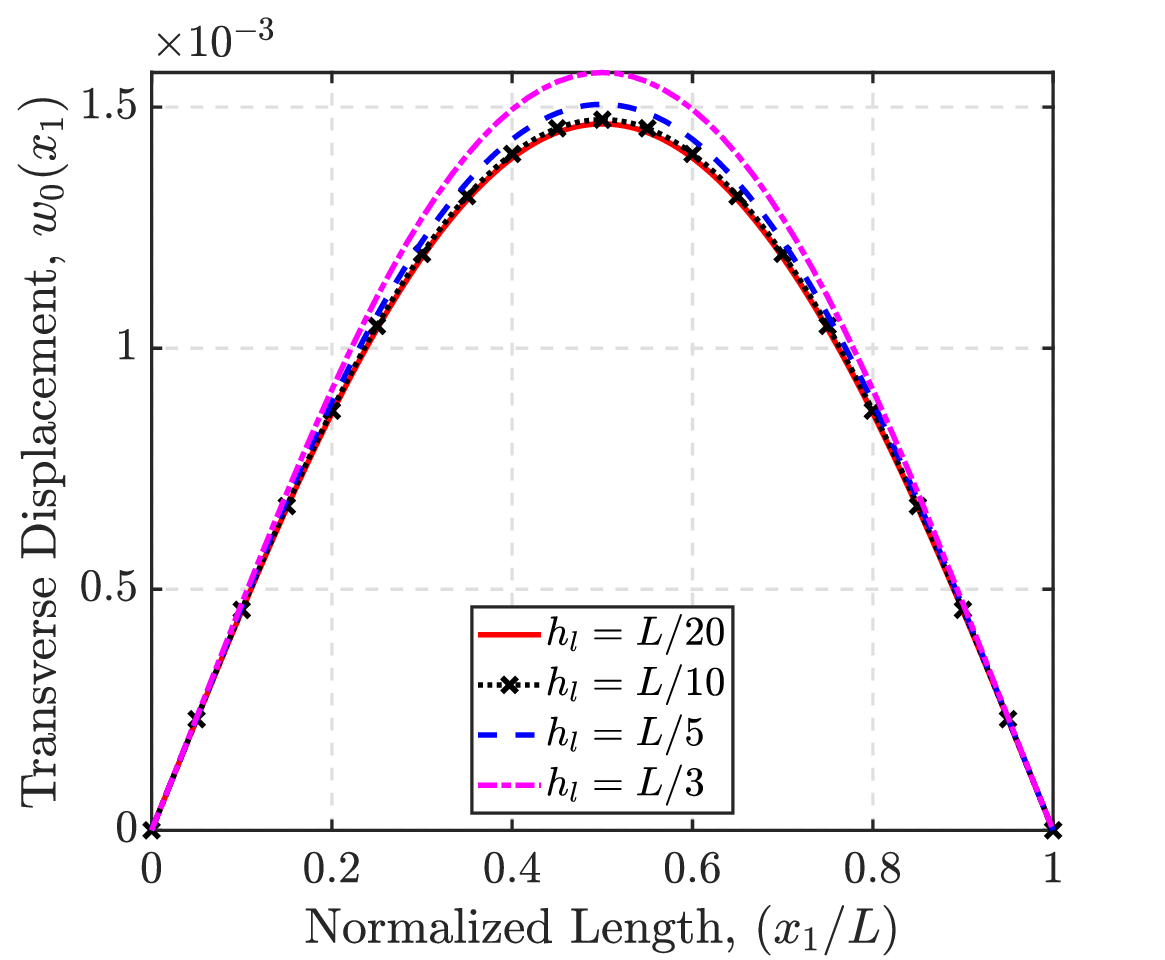}
	\caption{$w_0(x_1)$~vs~$h_l$ for $\alpha=0.8$}
	\label{fig:Result_3_SS_Actuation_WO_elec_load}
    \end{subfigure}
    \caption{Transverse displacement (in $\mathrm{m}$) of the simply supported smart beam for $q_0(x_1)=100~\mathrm{N/m}$ and $\phi=0~\mathrm{V}$.}
    \label{fig:Actution_simply h_l and alpha w/o elec load}
\end{figure}
First, a purely mechanical UDL of magnitude $q_0(x_1)=100~\mathrm{N/m}$ is applied over the entire smart beam. In the absence of electrical loads, the mechanical force through piezoelectric coupling is zero. Therefore, the smart structure behaves as an elastic structure. However, the elastic stiffness of the smart structure is modified due to the presence of the piezoelectric layer. The mid-plane transverse displacement $w_0(x_1)$ is compared for different values of the $\alpha$ and $h_l$ in Fig.~\ref{fig:Actution_simply h_l and alpha w/o elec load}.

In Fig.~\ref{fig:Result_2_SS_Actuation_WO_elec_load}, a softening in the elastic response of the smart structure is observed with the reduction of fractional-order $\alpha$. Similarly, an increase in the transverse displacement along the length of the beam is observed with an increasing horizon of nonlocal influence in Fig.~\ref{fig:Result_3_SS_Actuation_WO_elec_load}. Therefore, it is clear that the smart structure demonstrates a consistent softening (reduction in stiffness) with an increasing degree of nonlocal interactions. This observation is in agreement with the literature, where a consistent reduction in mechanical stiffness matrices is observed with increasing degree of nonlocal interactions\cite{patnaik2020ritz,sidhardh2021thermodynamics}. 

Subsequent to the above study, we apply only an electrical load to the smart structure ($q_0(x_1)=0$, $\phi_0(x_1)\neq 0$). The smart structure is mechanically actuated by the converse piezoelectric effect (nonzero $\{\mathbf{F_{ae}}\}$ and $\{\mathbf{F_{te}}\}$ in Eq.~\eqref{eq:element_stiffness_mat}). We note that the nonlocal interactions are realized over \textit{all} the system matrices: the stiffness and the mechanically induced forces due to the converse piezoelectric effect. To better highlight this, the transverse displacement of the nonlocal smart beam is evaluated when the piezoelectric layer is subject to an electric potential $\phi_0(x_1)=100$~V.

\begin{figure}[H]
    \centering
    \begin{subfigure}{.5\textwidth}
        \centering
        \includegraphics[width=1\linewidth]{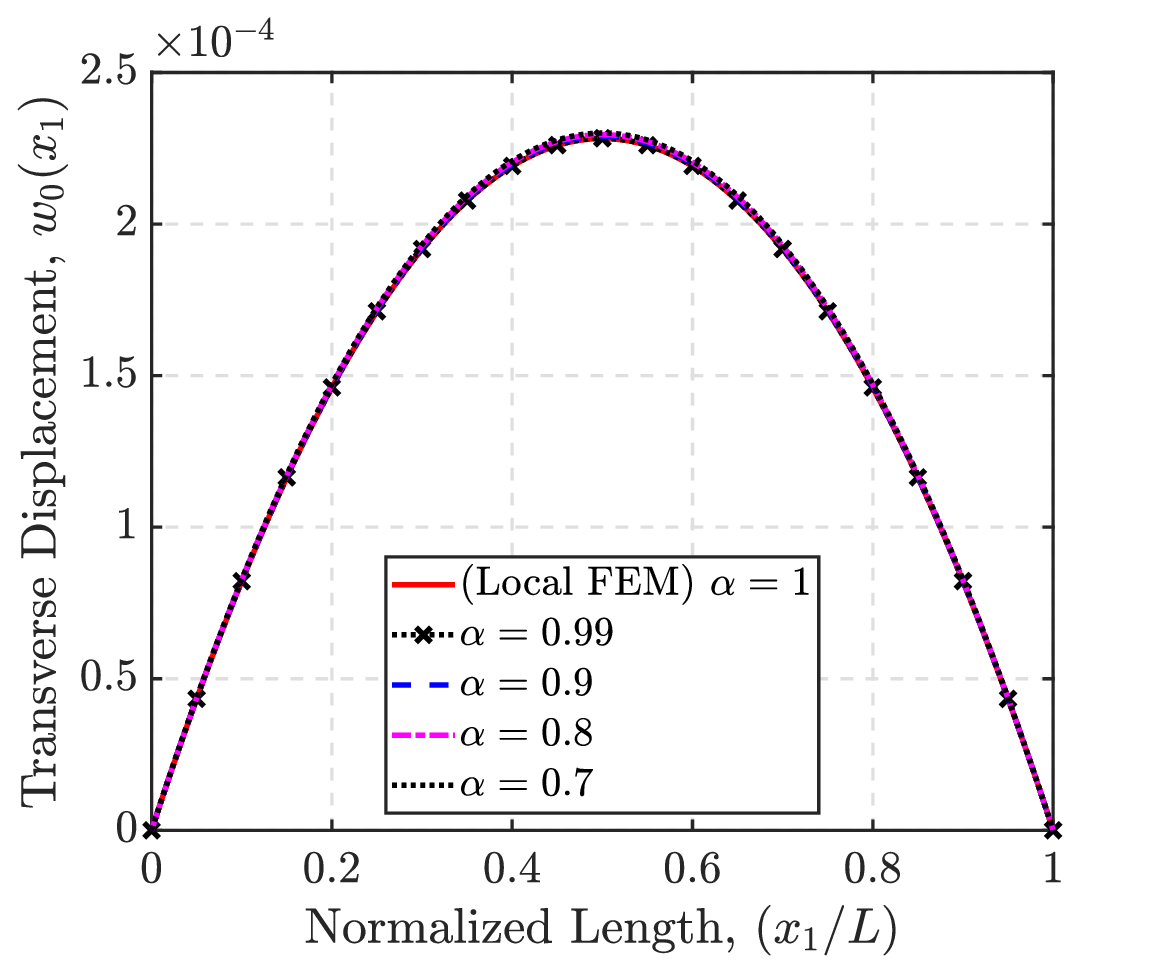}
        \caption{$w_0(x_1)$~vs~$\alpha$ for $h_l=L/5$}
        \label{fig:Result_2_SS_Actuation_WO_mech_load}
    \end{subfigure}%
    \begin{subfigure}{.5\textwidth}
	\centering
	\includegraphics[width=1\linewidth]{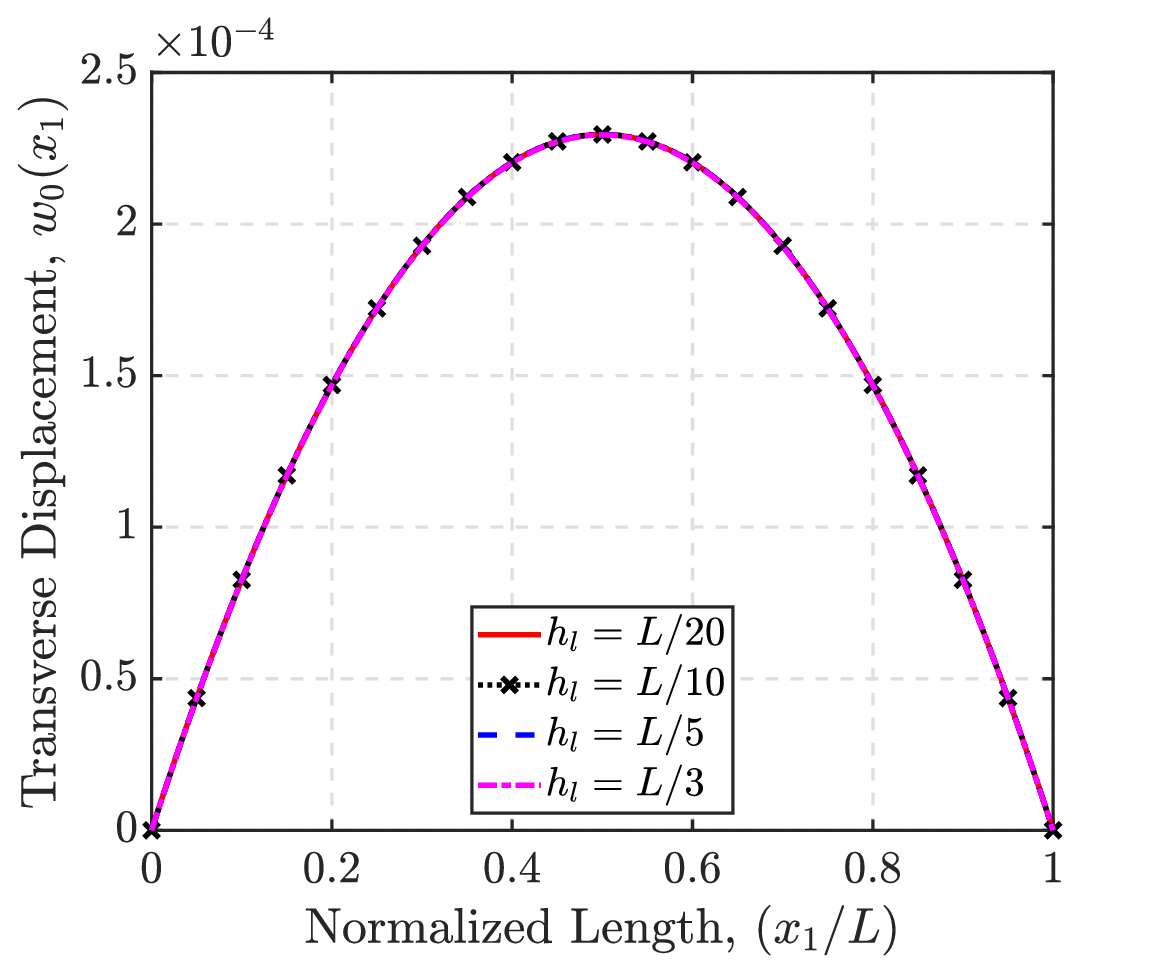}
	\caption{$w_0(x_1)$~vs~$h_l$ for $\alpha=0.8$}
	\label{fig:Result_3_SS_Actuation_WO_mech_load}
    \end{subfigure}
    \caption{Transverse displacement (in $\mathrm{m}$) of the simply supported smart beam for $q_0(x_1)=0~\mathrm{N/m}$ and $\phi=100~\mathrm{V}$.}
    \label{fig:Actution_simply h_l and alpha w/o mech load}
\end{figure}

The transverse displacement along the length of the beam for different numerical values of fractional-order constitutive parameters is compared in Fig.~\ref{fig:Actution_simply h_l and alpha w/o mech load}. Unlike earlier studies, no consistent softening (discernible) in elastic response is observed for increasing degree of nonlocal interactions. More clearly, no consistent increase in the maximum displacement is observed for reducing fractional-order or increasing horizon length of nonlocal influence. 

To better explain this, the system equations in Eq.~\eqref{eq:algebraic_gov_eq} are recast as follows:
\begin{subequations}
\label{eq: cons_discussion_matrix}
\begin{equation}
    [\mathbf{K}]\{\mathbf{X}\}=\{\mathbf{F}_m\}+\{\mathbf{F}_e\}
\end{equation}
where
\begin{equation}
\begin{split}
    [\mathbf{K}]=\begin{bmatrix}
         [\mathbf{K_{uu}}] & [\mathbf{K_{uw}}]\\
         [\mathbf{K_{uw}}]^T & [\mathbf{K_{ww}}]
    \end{bmatrix},~~\{\mathbf{X}\}^T=\begin{bmatrix}
        \{\mathbf{u}_g\}^T & \{\mathbf{w}_g\}^T
    \end{bmatrix},\\
    \{\mathbf{F}_m\}^T=\begin{bmatrix}
        \{\mathbf{F_{a}}\}^T & \{\mathbf{F_{t}}\}^T
    \end{bmatrix},~~\{\mathbf{F}_e\}^T=\begin{bmatrix}
        \{\mathbf{F_{ae}}\}^T & \{\mathbf{F_{te}}\}^T
    \end{bmatrix}
\end{split}
\end{equation}
\end{subequations}
In the previous case of a smart beam subject to purely mechanical load, it is established that the stiffness matrix $[\mathbf{K}]$ undergoes consistent reduction with an increase in the degree of nonlocal interactions. This is demonstrated as a consistent softening response for transverse displacement observed in Fig.~\ref{fig:Actution_simply h_l and alpha w/o elec load}. However, in the current study on the actuation of the smart beam via electrical load, the force vector $\{\mathbf{F}_e\}$ also depends on the degree of nonlocal interactions (see Eq.~\eqref{eq:elec_force_vector_u_w}). Therefore, an increase in the degree of nonlocal interactions results in a corresponding reduction in this vector, the mechanical force generated via the converse piezoelectric effect. This is clearly along the expected lines for the effect of nonlocal interactions on system matrices. Therefore, in the current study, both the stiffness $[\mathbf{K}]$ and the applied force vector $\{\mathbf{F}_e\}$ are inherently nonlocal in nature, which explains the variation of transverse displacements in Fig.~\ref{fig:Actution_simply h_l and alpha w/o mech load} with changing fractional-order constitutive parameters. This result is particularly interesting, given that, unlike previous studies on elastic response\cite{patnaik2020ritz,sidhardh2020geometrically} or one-way coupling\cite{sidhardh2021thermodynamics}, the force applied in this multiphysics coupling analysis is also related to nonlocal interactions. Therefore, displacement of the nonlocal structure is a net result of the combined softening of the system stiffness matrix and force vectors.

\begin{figure}[t]
    \centering
    \begin{subfigure}{.5\textwidth}
        \centering
        \includegraphics[width=1\linewidth]{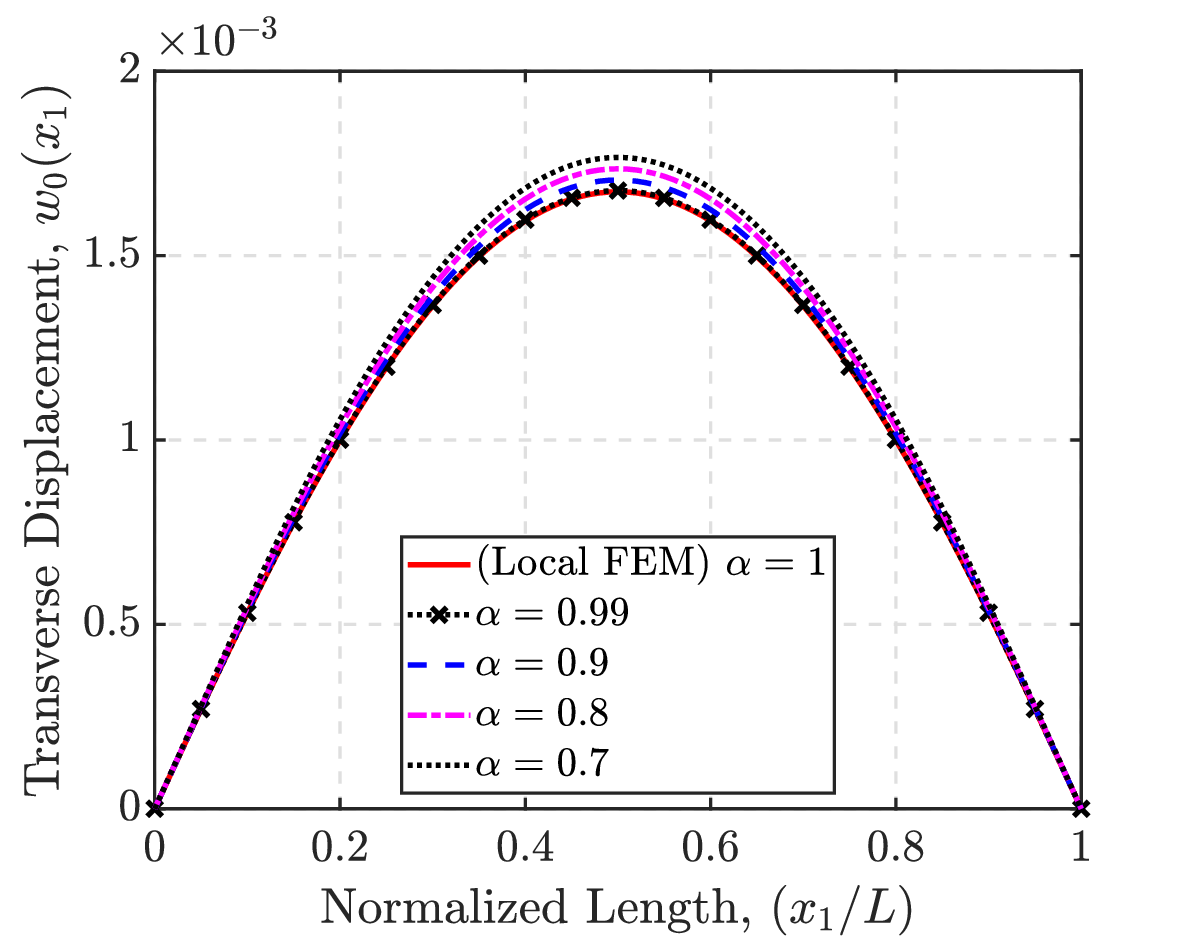}
        \caption{$w_0(x_1)$~vs~$\alpha$ for $h_l=L/5$}
        \label{fig:Result_2_SS_Actuation}
    \end{subfigure}%
    \begin{subfigure}{.5\textwidth}
	\centering
	\includegraphics[width=1\linewidth]{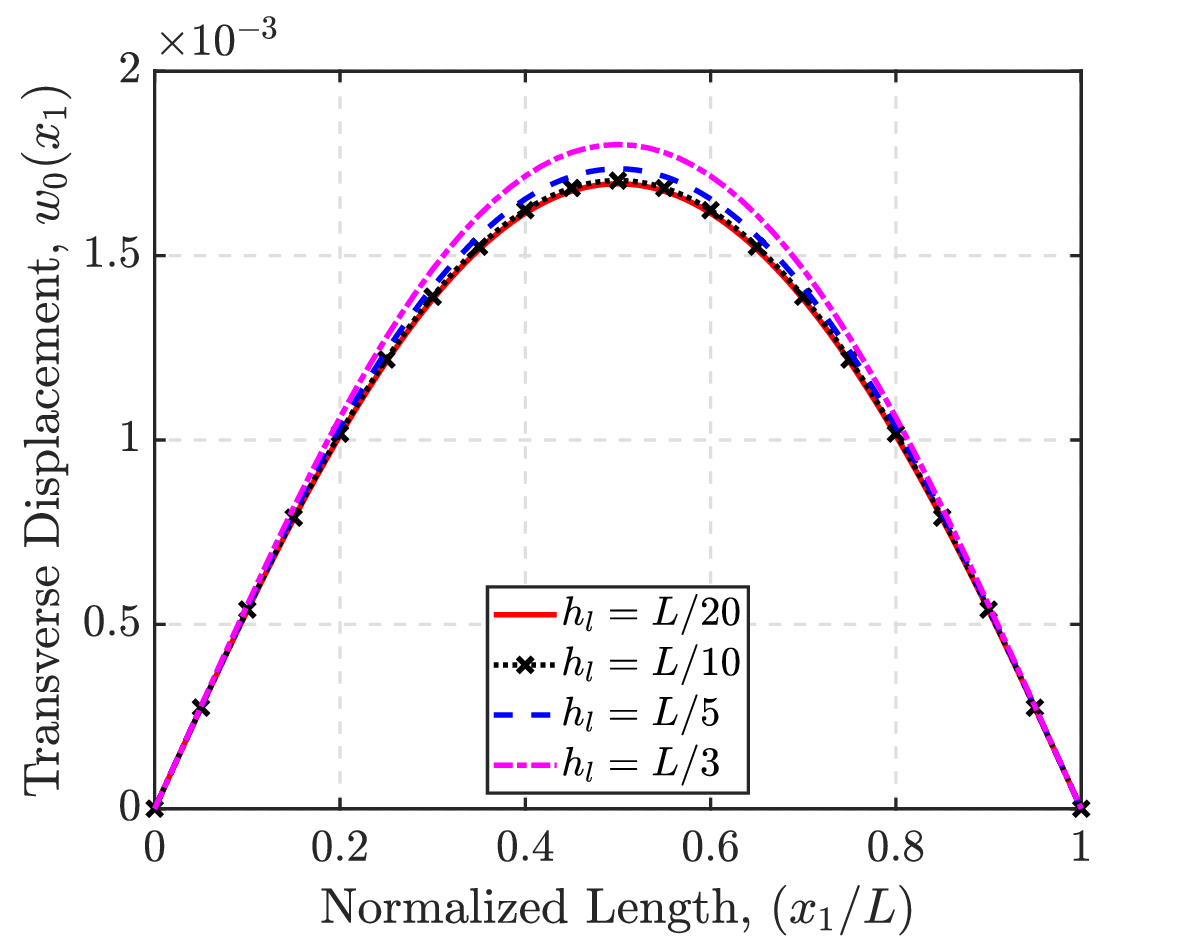}
	\caption{$w_0(x_1)$~vs~$h_l$ for $\alpha=0.8$}
	\label{fig:Result_3_SS_Actuation}
    \end{subfigure}
    \caption{Transverse displacement (in $\mathrm{m}$) of the simply supported smart beam for $q_0(x_1)=100~\mathrm{N/m}$ and $\phi=100~\mathrm{V}$.}
    \label{fig:Actution_simply h_l and alpha}
\end{figure}

Finally, we study the elastic response of the smart beam when subject to a combined electro-mechanical load. In this case, the smart beam is subject to a mechanical load of $q_0(x_1)=100~\mathrm{N/m}$ and an electrical load of $\phi_0(x_1)=100~\mathrm{V}$, simultaneously. For different fractional-order constitutive parameters $\alpha$ and $h_l$, the transverse displacement of the smart beam is solved and is presented in Fig.~\ref{fig:Actution_simply h_l and alpha}. A consistent softening of the smart structure is observed with either reducing fractional-order $\alpha$ in Fig.~\ref{fig:Result_2_SS_Actuation}, or increasing the horizon of nonlocal influence in Fig.~\ref{fig:Result_3_SS_Actuation}. This observation clearly follows from the detailed discussion presented earlier.
   
\paragraph{\textit{Direct piezoelectric effect}}~\\
\noindent
In this study, we propose to analyze the effect of nonlocal interactions on direct piezoelectric coupling. More clearly, we study the electrical potential generated across the piezoelectric layer when a simply-supported nonlocal smart beam is subject to a uniformly distributed mechanical load $q_0(x_1)=1~\mathrm{N/m}$. The smart structure is configured in an open-circuit arrangement, where the bottom surface of the piezoelectric layer/patch attached to the substrate is connected to the ground ($\phi(x_1,h/2)=0$ in Eq.~\eqref{eq: elec_potential}) and the electrical potential on the top surface is a free variable. This configuration generates an electrical potential difference across the piezoelectric material in response to the applied external mechanical load. 

\begin{figure} [H]
    \centering
    \begin{subfigure}{.5\textwidth}
	\centering
	\includegraphics[width=1\linewidth]{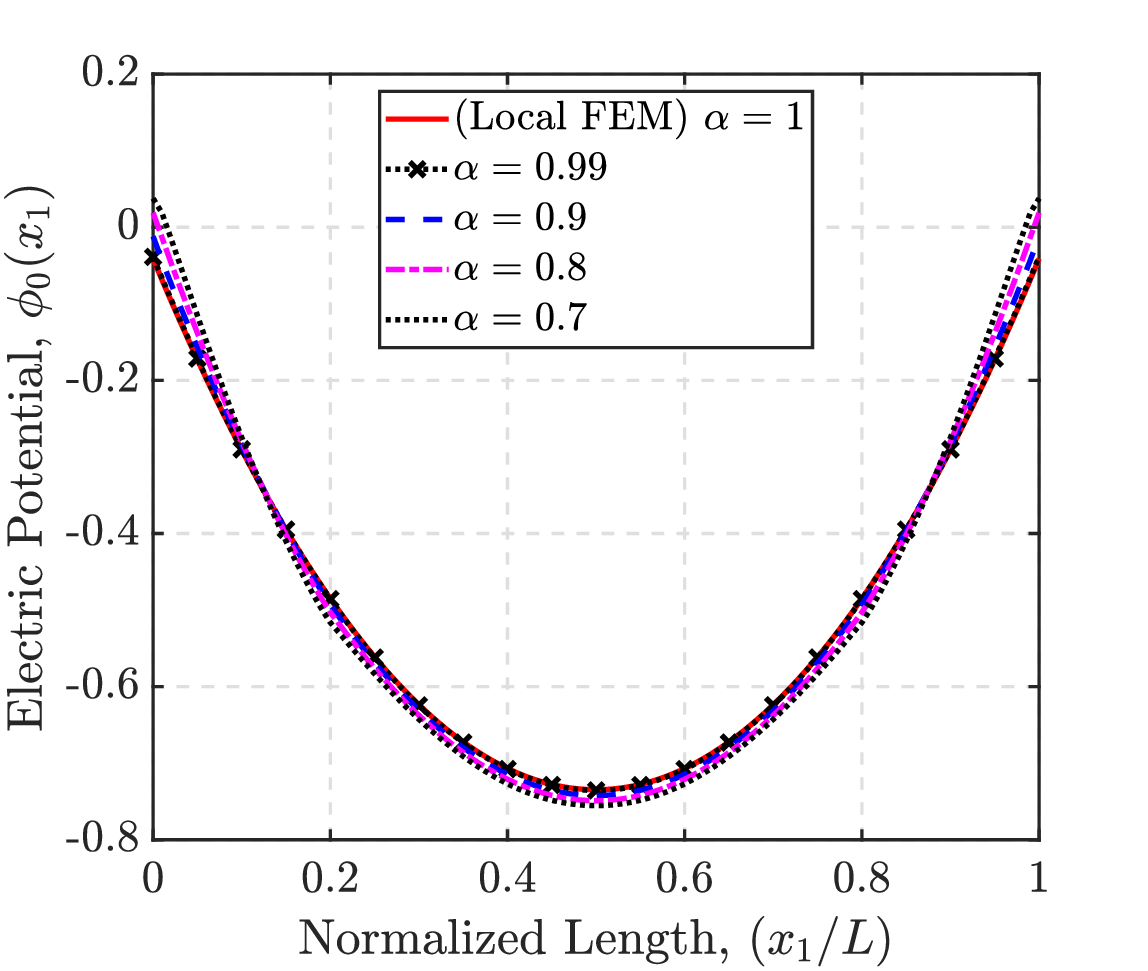}
	\caption{$\phi$~vs~$\alpha$ for $h_l=L/5$}
	\label{fig:Result_2_SS_Sansing}
    \end{subfigure}%
    \begin{subfigure}{.5\textwidth}
	\centering
	\includegraphics[width=1\linewidth]{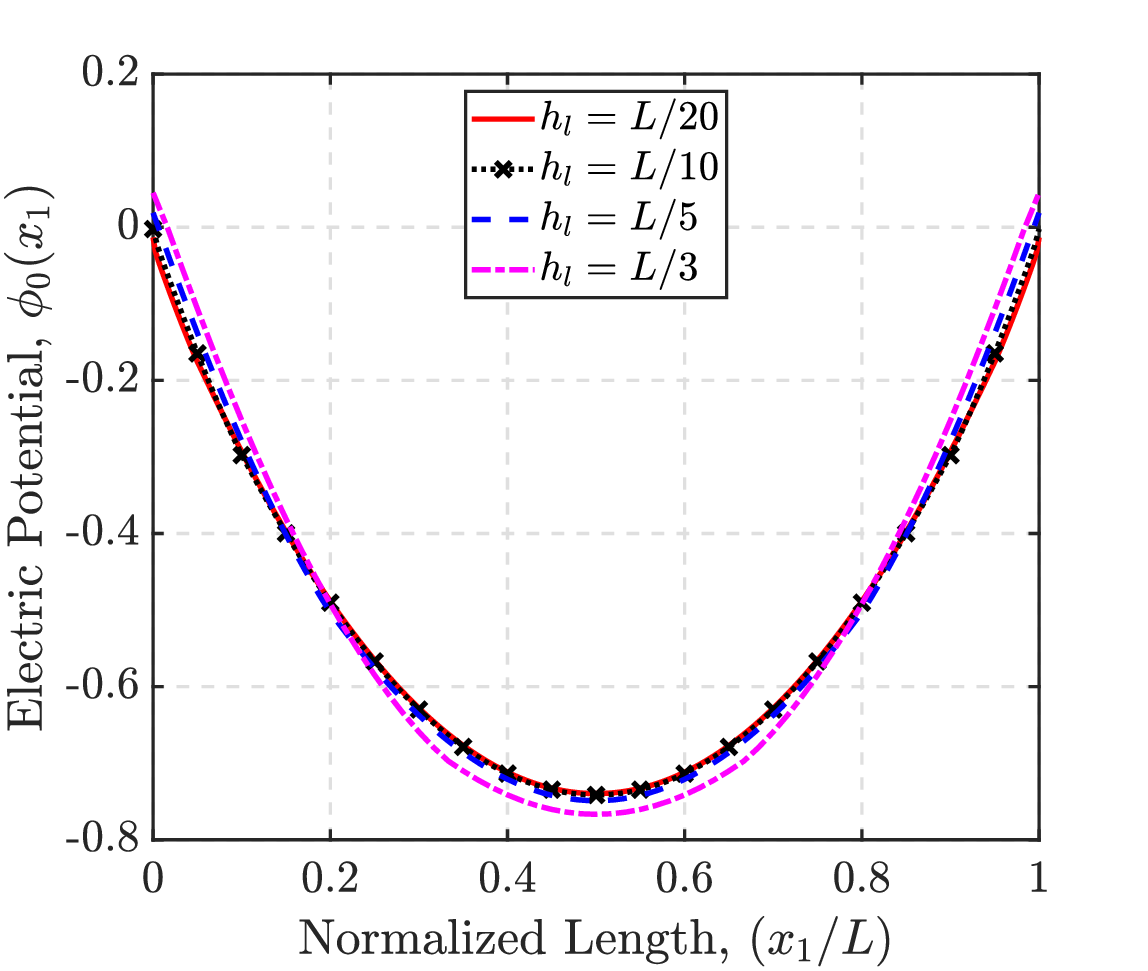}
	\caption{$\phi$~vs~$h_l$ for $\alpha=0.8$}
	\label{fig:Result_3_SS_Sensing}
    \end{subfigure}
    \caption{Electrical potential (in $\mathrm{V}$) of the simply supported smart beam for $q_0(x_1)=1~\mathrm{N/m}$.}
    \label{fig:Sensing_simply h_l and alpha}
\end{figure}

The electrical potential generated across the length of the simply supported beam for different fractional-order constitutive parameters is provided in Fig.~\ref{fig:Sensing_simply h_l and alpha}. More specifically, the electrical potential generated at the top surface of the piezoelectric layer for varying fractional-order and horizon of nonlocal influence is provided in Figs.~\ref{fig:Result_2_SS_Sansing} and \ref{fig:Result_3_SS_Sensing}, respectively. From these figures, it is clear that the electrical potential generated at the top surface of the piezoelectric layer varies with changing fractional-order constitutive parameters. However, the trend for variation of electrical energy harvesting with a degree of nonlocal interactions is unclear from these figures. Therefore, for ease of comparison, a different metric in root mean square (RMS) voltage is proposed to quantify the electrical energy harvested. The scalar RMS voltage is defined as\cite{j2004electric}:
\begin{equation}
\label{eq: rms}
   V_{rms}=\sqrt{\frac{\{\bm{\phi}_g\}^T\{\bm{\phi}_g\}}{\mathcal{N_P}+1}}
\end{equation}

\begin{table}[H]
\begin{center}
    \begin{tabular}{ c c c c c } 
    \hline
    \hline
    $h_l$ & \multicolumn{4}{c}{$V_{rms}$~(in $\mathrm{V}$)} \\ 
    \hline
     & $\alpha=1.0$ & $\alpha=0.9$ & $\alpha=0.8$ & $\alpha=0.7$ \\
     \hline
    $L/20$ & 0.4512 & 0.4526 & 0.4538 & 0.4548 \\ [1.5ex]   
     $L/10$ & 0.4512 & 0.4527 & 0.4539 & 0.4550 \\[1.5ex]
     $L/5$ & 0.4512 & 0.4543 & 0.4573 & 0.4606 \\   
    \hline
    \hline
    \end{tabular}
    \caption{RMS voltage generated across the thickness of the piezoelectric layer of the simply supported smart beam for $q_0(x_1)=1~\mathrm{N/m}$.}
    \label{tab:RMS voltage layer}
\end{center}
\end{table}
Recall that $\{\bm{\phi}_g\}$ and $\mathcal{N_P}$ are defined as the vector for nodal values of electrical potential and number of elements spanning over piezoelectric layer in the f-FE model, respectively. Physically, the scalar RMS voltage is representative of the electrical energy harvested from the smart beam via direct piezoelectric coupling. The RMS voltage generated by the smart beam for different fractional-order constitutive parameters is presented in Table~\ref{tab:RMS voltage layer}. Note that in this study, we choose $\mathcal{N}=500$, in keeping with the dynamic rate of convergence for all choices of fractional-order constitutive parameters.

From the table, a consistent increase is observed in the RMS voltage generated across the piezoelectric layer with an increasing degree of nonlocal interactions. More clearly, the RMS voltage increases with a reduction of fractional-order $\alpha$ or with an increase in the horizon of nonlocal influence $h_l$. This points to a consistent increase in the piezoelectric coupling with increasing nonlocal interactions and, therefore, presents interesting possibilities towards enhancing piezoelectric coupling by tuning the nonlocal interactions. This observation of an increase in the electrical potential with increasing nonlocal interactions is akin to greater electrical energy harvested experimentally from substrate beams of complex geometry\cite{zhao2014broadband,zhao2015experimental,matova2013effect}. Recall that, it is established in literature, such structures with complex geometries and/or material distributions presenting a multiscale architecture can be successfully modeled as fractional-order elastic models\cite{patnaik2022variable,patnaik2020generalized}. Therefore, the observation of an increment in electrical energy due to nonlocal interactions (varying $\alpha$ and $h_l$) may be corroborated by above experimental studies.\\

\noindent
\textbf{Smart beam with non-neglected electrodes:} The assumption to ignore the electrodes within the numerical model developed here is in keeping with most literature\cite{erturk2009experimentally,erturk2008distributed,wu2013novel}. This is because of the extremely small dimensions (thickness) of the electrode with respect to the piezoelectric layer ($<h_{P}/100$), rendering the contribution of the electrodes to the mechanical stiffness negligible. For the sake of completeness, we present here a case study for non-neglected electrodes to realise their effect on the energy harvested. For this purpose, we have modified our numerical model to include thin-film electrodes\cite{piezoelement} in Parallel Plate Electrode (PPE) configuration over the nonlocal smart structure.  More specifically, we consider integer-order constitutive relations for the electrode over the fractional-order model for smart structure. The electrodes are considered to be made of Aluminium\cite{reed1990physical} with material modulus: $E=68~\mathrm{GPa}$. The thickness of the electrode is chosen to be $h_e=5\times10^{-6}~\mathrm{m}$ as given in literature \cite{reed1990physical,suchanek1982influence,xu2019modeling,piezoelement}. The material and geometrical properties of smart structure in this study are identical to those considered previously, with the thickness of the piezoelectric layer $h_P=0.265~\mathrm{mm}$ as considered in the experimental studies by Erturk and Inman\cite{erturk2011piezoelectric} and later studied analytically in \cite{wang2013analysis}.  We repeat the numerical simulations for the smart beam with a piezoelectric layer under a uniformly distributed mechanical load of 1~$\mathrm{N/m}$ subject to simply supported boundary conditions. This is a trivial extension of the numerical model developed here, hence complete details of modifications in existing model are skipped here for the sake of brevity. The RMS voltages for the smart beam with and without electrodes are compared for the local and nonlocal elastic case in Table \ref{tab:RMS voltage layer with electrode}.
\noindent
\begin{table}[H]
\begin{center}
    \begin{tabular}{ c | c c c c c} 
    \hline
    \hline
    & \multicolumn{5}{c}{$V_{rms}$~(in $\mathrm{V})$} \\ 
    \hline
     & Local & $\alpha=0.99$ & $\alpha=0.9$ & $\alpha=0.8$ & $\alpha=0.7$ \\[1.5ex]
     \hline
    Without electrodes & $0.4341$ & $0.4343$ & $0.4362$ & $0.4380$ & $0.4397$\\ [1.5ex]   
     With electrode & $0.4053$ & $0.4056$ & $0.4073$ & $0.4090$ & $0.4106$\\[1.5ex]
     Difference (in \%) & 6.63 & 6.60 & 6.62 & 6.62 &  6.62 \\   
    \hline
    \hline
    \end{tabular}
    \caption{Comparison  of electric potential corresponding to smart structure with and without electrodes in the numerical model.}
    \label{tab:RMS voltage layer with electrode}
\end{center}
\end{table}
Based on the above findings, it is clear that the effect of electrode over the electrical energy harvested is realized in increasing the mechanical stiffness of the smart structure. Moreover, this effect is minimal and uniform ($6.6\%$) for local and nonlocal models. More clearly, the difference in ${V_{rms}}$ voltage harvested is almost identical for local and nonlocal models. Therefore, we neglect the electrodes in the subsequent studies.

\subsubsection{Smart beam with piezoelectric patch}
In this case study, we propose to study the effect of nonlocal interactions over the piezoelectric response of a nonlocal smart beam with a piezoelectric patch. Unlike the previous case, the current case presents differential nonlocal interactions within the substrate and the piezoelectric patch. This is in contrast to the previous case-study, where we assume that the length scales for the piezoelectric layer and the substrate beam are identical at each point along the length of the smart beam. Instead,  we depart from this assumption by a differential truncation of the nonlocal length scales within the piezoelectric patch. An illustration of the same is provided in Figure \ref{fig:nonlocal_patch_intrection}. Therefore, in this study, we treat the nonlocal length scales separately for the piezoelectric patch and the substrate beam. Further, note that the previous study on smart beam with a piezoelectric layer can be considered as a specific case of this case study.

In this study, we consider a smart cantilever beam with a piezoelectric patch placed such that $x_0=0$ and $L_{p}=0.3L$, where $L$ is the total length of the substrate. This location for the piezoelectric patch is chosen to maximize the piezoelectric coupling within the smart beam\cite{xie2014potential}. As done previously, we investigate the effect of mechanical, electrical and combined loads over the electro-mechanical response of the smart beam.

\paragraph{\textit{Converse piezoelectric effect}}~\\
\begin{figure} [H]
    \centering
    \begin{subfigure}{.5\textwidth}
	\centering
	\includegraphics[width=1\linewidth]{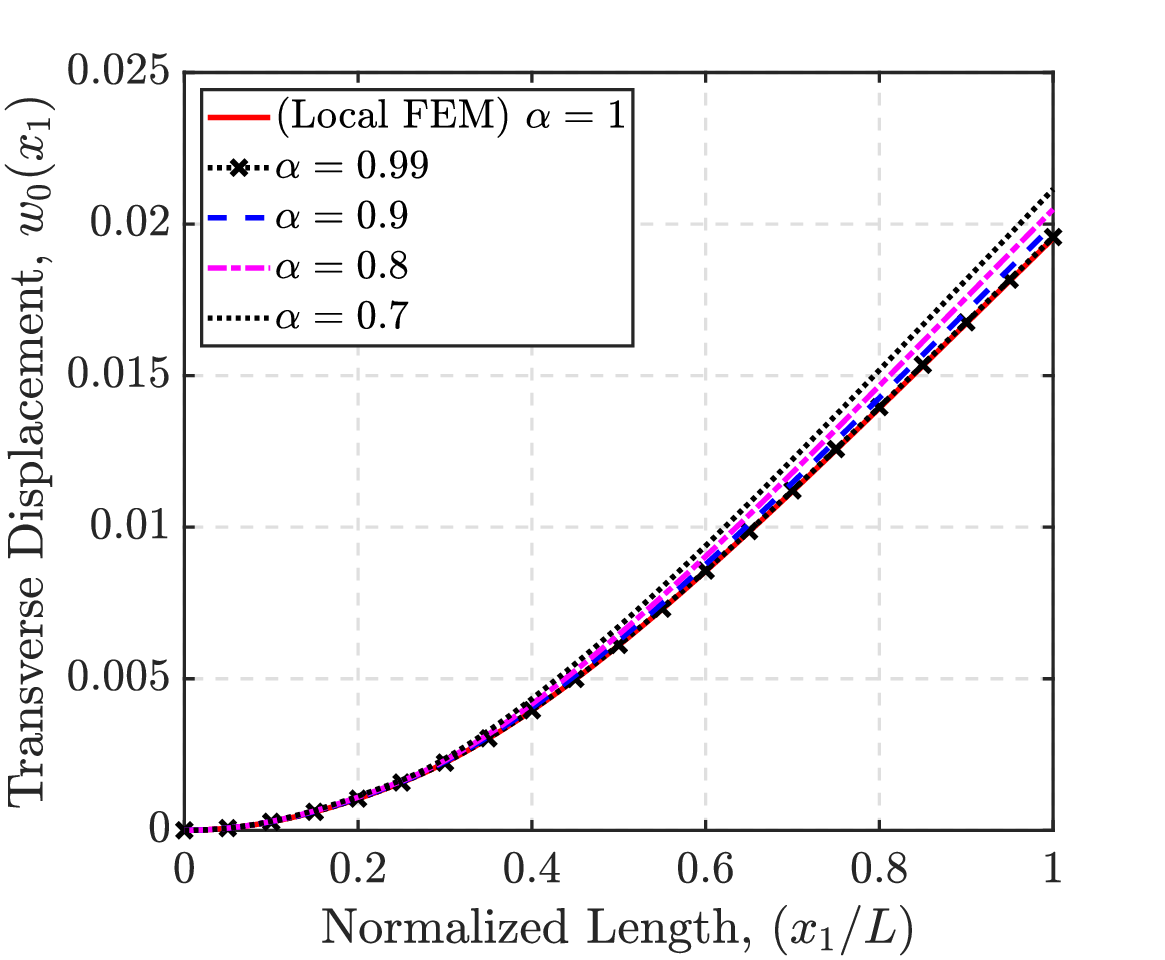}
	\caption{$w_0(x_1)$~vs~$\alpha$ for $h_l=L/5$}
	\label{fig:Result_2_CF_Patch_varing_alpha_without_V}
    \end{subfigure}%
    \begin{subfigure}{.5\textwidth}
	\centering
	\includegraphics[width=1\linewidth]{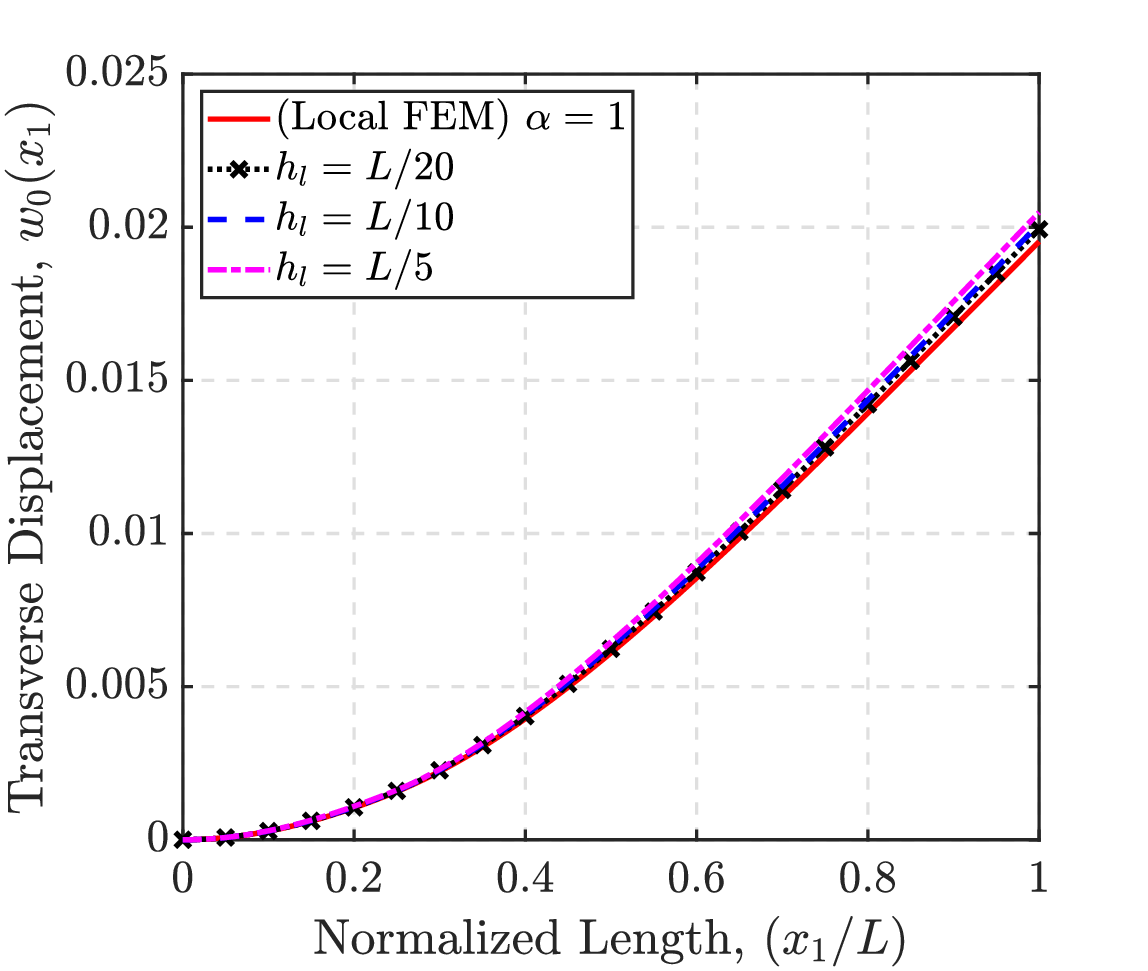}
	\caption{$w_0(x_1)$~vs~$h_l$ for $\alpha=0.8$}
	\label{fig:Result_3_CF_Patch_varing_hl_without_V}
    \end{subfigure}
    \caption{Transverse displacement (in $\mathrm{m}$) of the cantilever smart beam subjected to $q_0(x_1)=100~\mathrm{N/m}$ and $\phi=0~\mathrm{V}$.}
    \label{fig:patch_CF h_l and alpha without V}
\end{figure}
In this study, the governing equations developed in Eq.~\eqref{eq:algebraic_gov_eq} are solved for the mechanical response of the smart beam with the piezoelectric patch demonstrating the converse piezoelectric effect. We consider the following three different cases of electro-mechanical loading: (i) a uniformly distributed mechanical load $q_0(x_1)=100~\mathrm{N/m}$ along the length of the substrate; (ii) a uniform electrical potential $\phi_0(x_1)=50~\mathrm{V}$ along the length of the piezoelectric patch; (iii) a combination of the above electro-mechanical loads. 

The transverse displacement along the length of the smart beam when subjected to purely mechanical loads (Case-i) for different fractional-order constitutive parameters is compared in Fig.~\ref{fig:patch_CF h_l and alpha without V}. Clearly, a consistent reduction in stiffness, evident from the increase in deformation, is observed with an increase in the degree of nonlocal interactions. This points to nonlocal interactions over the elastic response that presents a consistent softening of the structural stiffness as reported in the literature\cite{patnaik2020ritz,sidhardh2021thermodynamics}.

\begin{figure} [H]
    \centering
    \begin{subfigure}{.5\textwidth}
	\centering
	\includegraphics[width=1\linewidth]{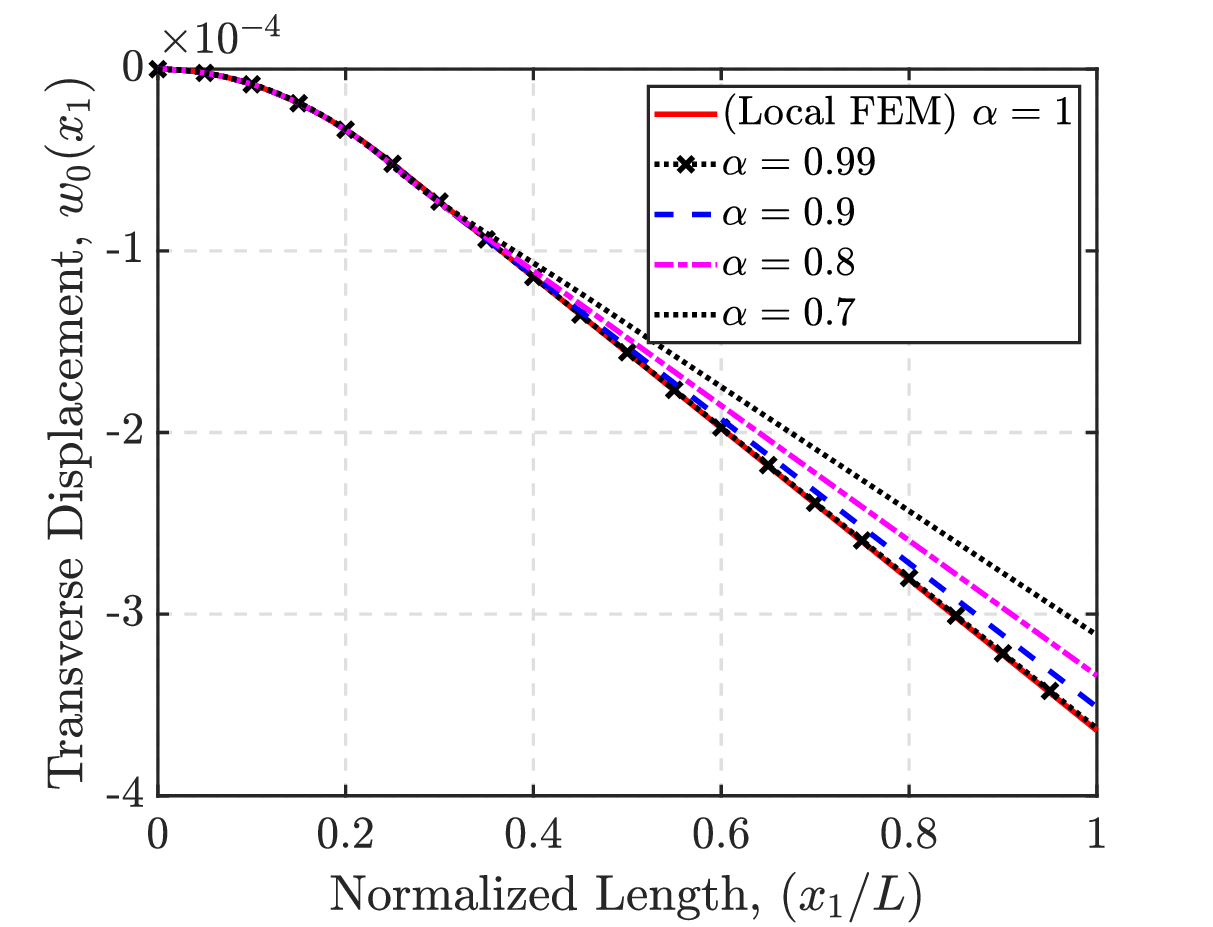}
	\caption{$w_0(x_1)$~vs~$\alpha$ for $h_l=L/5$}
	\label{fig:Result_2_CF_Patch_varing_alpha_without_mech_load}
    \end{subfigure}%
    \begin{subfigure}{.5\textwidth}
	\centering
	\includegraphics[width=1\linewidth]{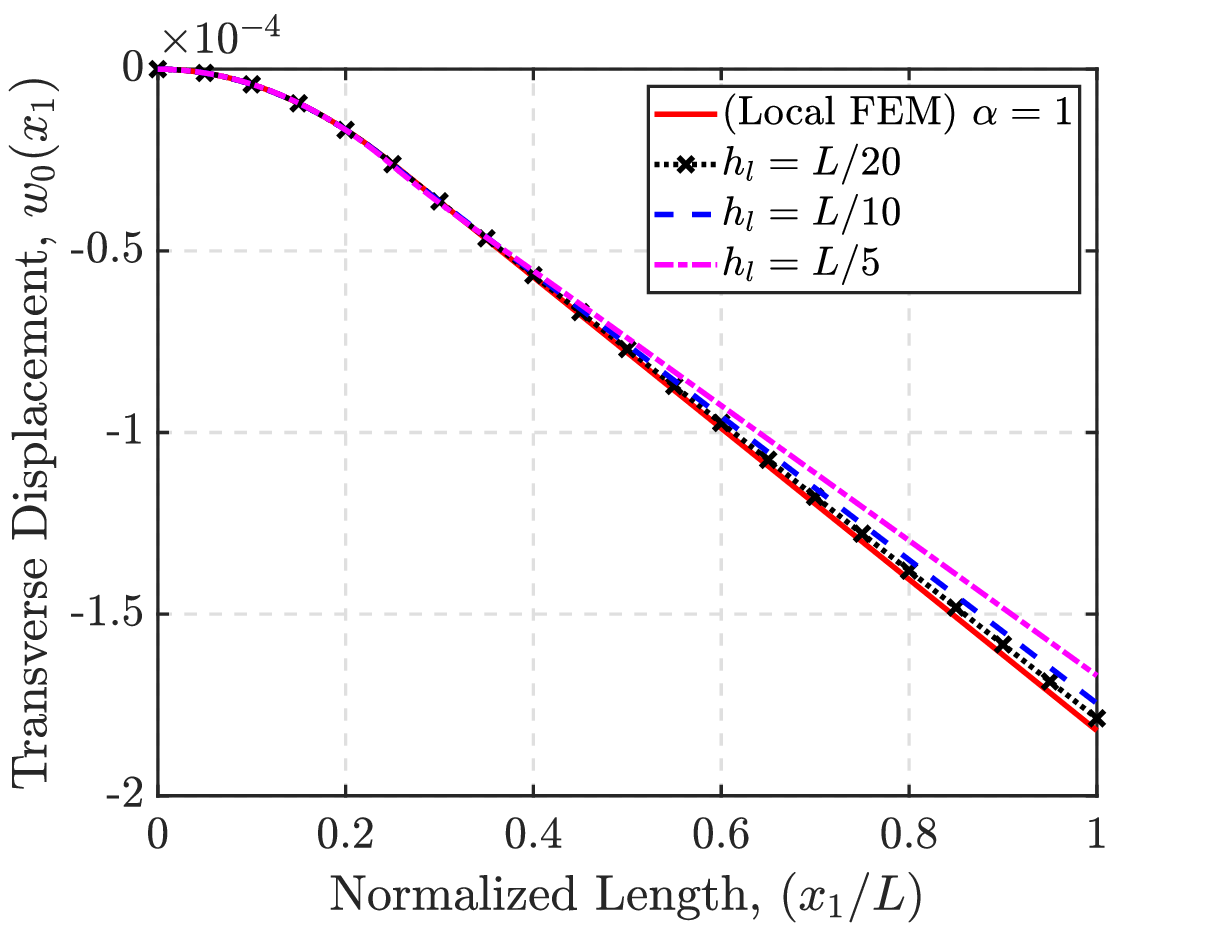}
	\caption{$w_0(x_1)$~vs~$h_l$ for $\alpha=0.8$}
	\label{fig:Result_3_CF_Patch_varing_hl_without_mech_load}
    \end{subfigure}
    \caption{Transverse displacement (in $\mathrm{m}$) of the cantilever smart beam subjected to $q_0(x_1)=0~\mathrm{N/m}$ and $\phi=50~\mathrm{V}$.}
    \label{fig:patch_CF h_l and alpha without mech load}
\end{figure}

Next, we subject the piezoelectric patch to uniformly distributed electrical loads, resulting in a mechanical actuation of the substrate (Case-ii). The transverse displacement along the length of the smart beam for purely electrical load is compared for different fractional-order constitutive parameters in Fig.~\ref{fig:patch_CF h_l and alpha without mech load}. In this case, for an identical electrical load, the transverse displacement of the smart beam consistently reduces (magnitude) with an increase in the degree of nonlocal interactions. This is contrary to the softening (increasing displacement) observed earlier with an increasing degree of nonlocality (Case-i). However, recall that, unlike previous studies, the mechanical force caused in response to electrical load via converse piezoelectric effect also undergoes reduction within an increase in the degree of nonlocal interactions.

This is similar to the result in Fig.~\ref{fig:Actution_simply h_l and alpha w/o mech load} discussed earlier. Unlike the simply supported smart beam studied earlier, here the reduction in induced mechanical force outweighs the corresponding softening in mechanical stiffness for the cantilever beam. 
\begin{figure} [H]
    \centering
    \begin{subfigure}{.5\textwidth}
	\centering
	\includegraphics[width=1\linewidth]{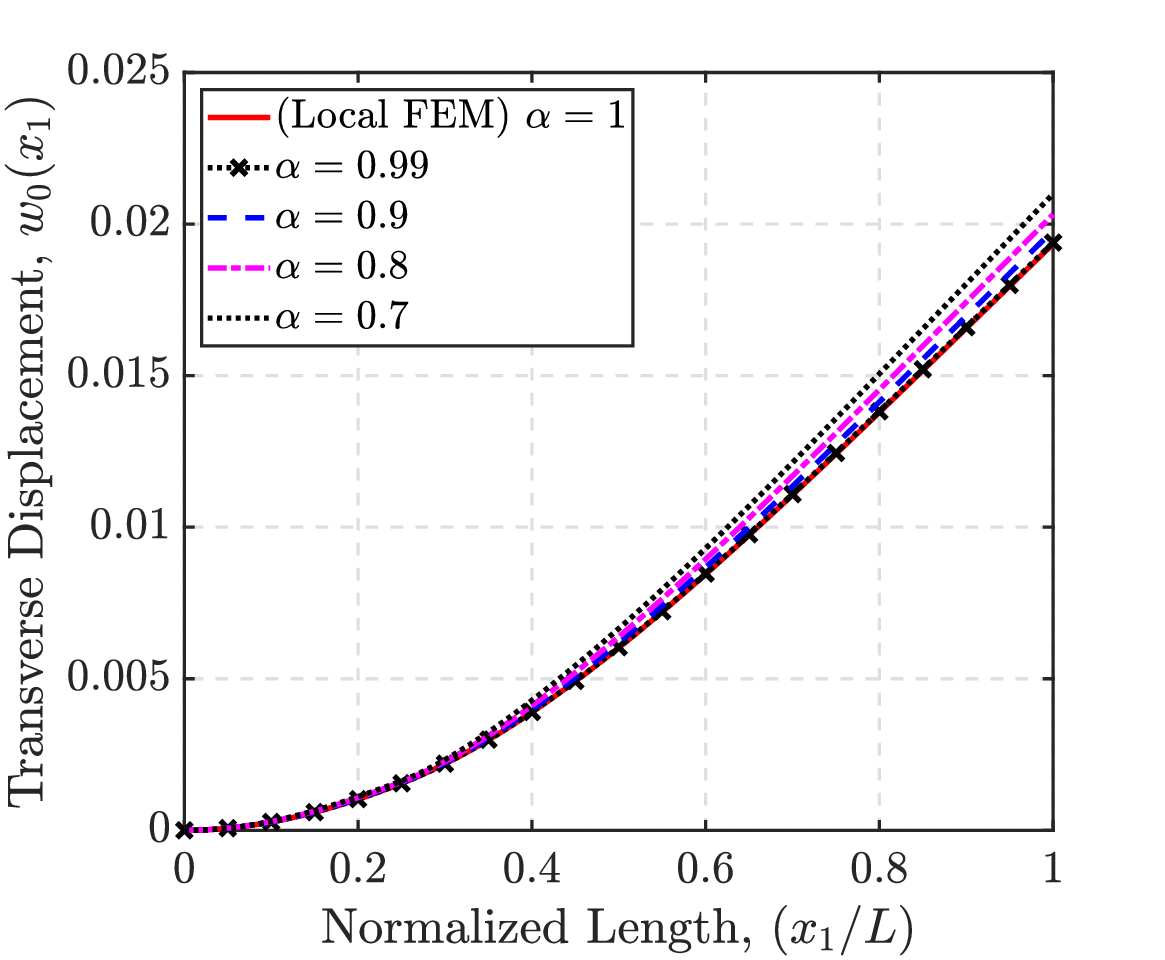}
	\caption{$w_0(x_1)$~vs~$\alpha$ for $h_l=L/5$}
	\label{fig:Result_2_CF_Patch_varing_alpha_combined_load}
    \end{subfigure}%
    \begin{subfigure}{.5\textwidth}
	\centering
	\includegraphics[width=1\linewidth]{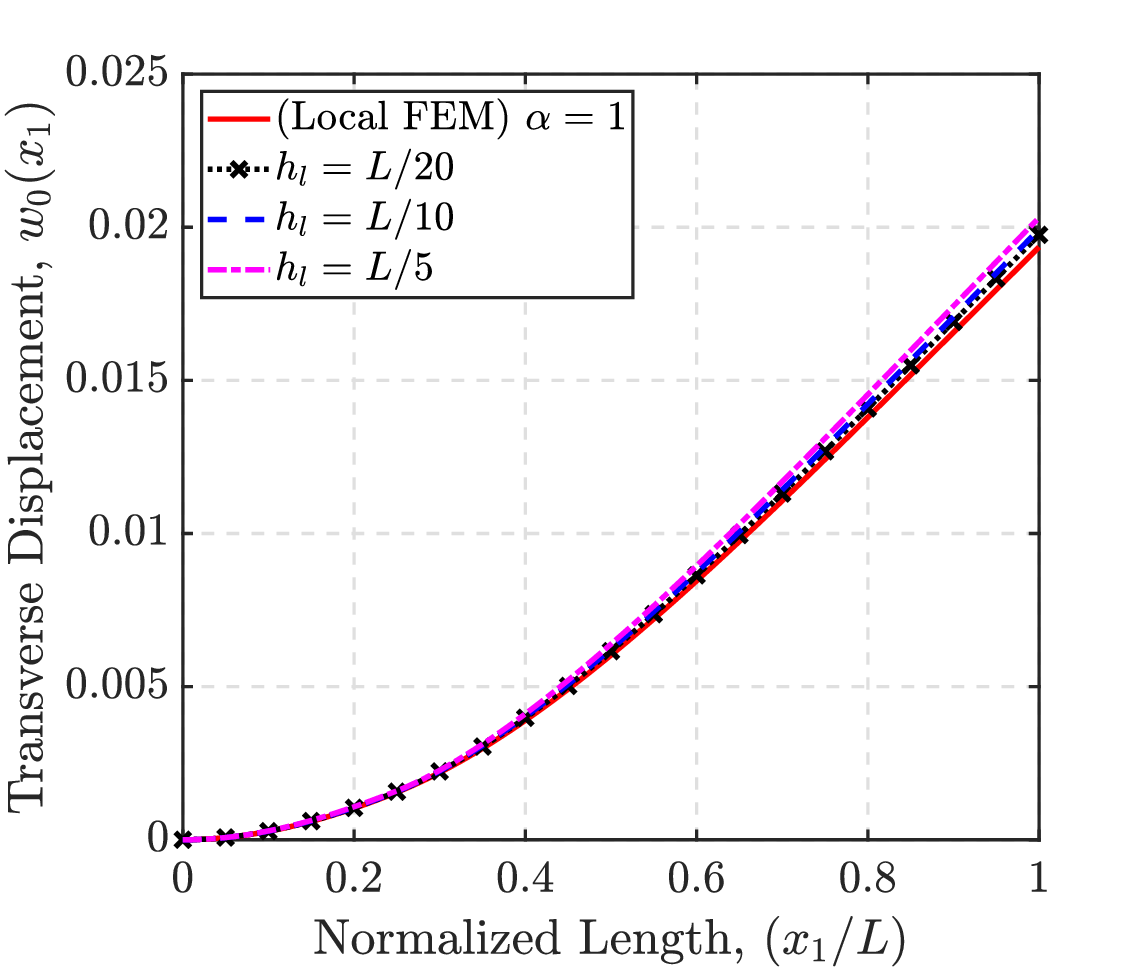}
	\caption{$w_0(x_1)$~vs~$h_l$ for $\alpha=0.8$}
	\label{fig:Result_3_CF_Patch_varing_hl_combined_load}
    \end{subfigure}
    \caption{Transverse displacement (in $\mathrm{m}$) of the cantilever smart beam subjected to $q_0(x_1)=100~\mathrm{N/m}$ and $\phi=50~\mathrm{V}$.}
    \label{fig:patch_CF h_l and alpha combined load}
\end{figure}

Finally, the transverse displacement of the smart beam for a combined electro-mechanical load is provided in Fig.~\ref{fig:patch_CF h_l and alpha combined load}. In this case, the deformation of the smart beam for different fractional-order constitutive parameters is a linear superposition of the individual responses to mechanical and electrical loads. Therefore, the effect of varying fractional-order constitutive parameters also is a net result of their effect on individual studies carried out in the above studies. Along these lines, we merely note that for the electro-mechanical loads considered here, the influence of the electric potential is relatively weak in comparison to that of the mechanical load.

\paragraph{\textit{Direct piezoelectric effect}}~\\
\noindent
Finally, we investigate the effect of nonlocal interactions over the direct piezoelectric coupling within a cantilever smart beam with a piezoelectric patch. The smart structure is subject to a uniformly distributed load $q_0(x_1)=1~\mathrm{N/m}$ along the length of the substrate. The piezoelectric patch is configured in an open-circuit arrangement, where the bottom surface is connected to the ground while the top surface remains free. Due to the application of an external load in this setup, an electric field is generated within the piezoelectric patch. We solve the algebraic governing equations in Eq.~\eqref{eq:algebraic_gov_eq_sensing} for different fractional-order parameters and thereby realize the influence of the degree of nonlocal interactions over the electrical potential generated at the top surface of the piezoelectric patch. 

\begin{figure} [H]
    \centering
    \begin{subfigure}{.5\textwidth}
        \centering
        \includegraphics[width=1\linewidth]{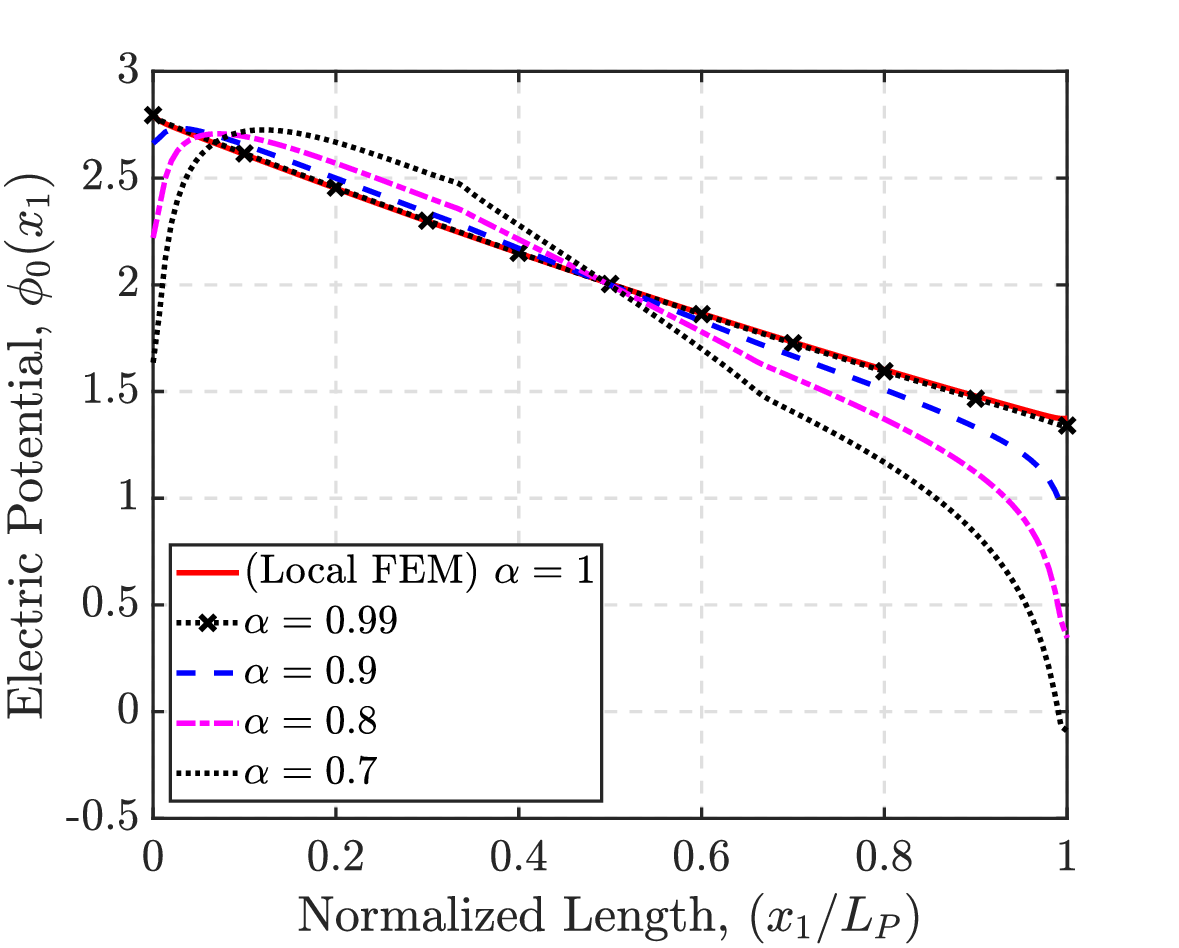}
        \caption{$\phi$~vs~$\alpha$ for $h_l=L/5$}
        \label{fig:Result_1_CF_Sansing}
    \end{subfigure}%
    \begin{subfigure}{.5\textwidth}
        \centering
        \includegraphics[width=1\linewidth]{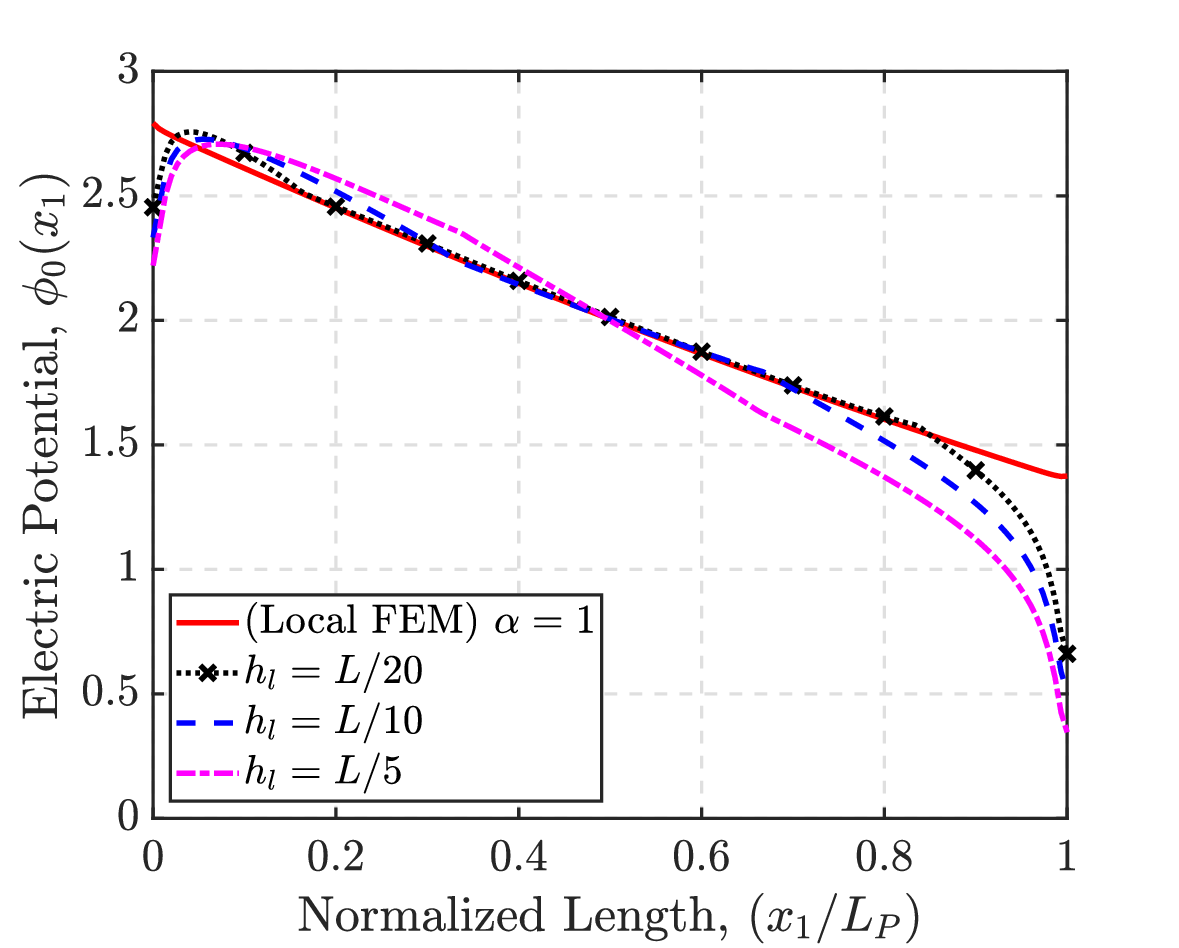}
        \caption{$\phi$~vs~$h_l$ for $\alpha=0.8$}
        \label{fig:Result_2_CF_Sensing}
    \end{subfigure}
    \caption{Electrical potential (in $\mathrm{V}$) of the cantilever smart beam for $q_0(x_1)=1~\mathrm{N/m}$.}
    \label{fig:Sensing_CF h_l and alpha}
\end{figure}

The electrical potential across the length of the piezoelectric patch ($x_1=0, L_P$) is compared for different fractional-order constitutive parameters in Fig.~\ref{fig:Sensing_CF h_l and alpha}. Clearly, a change in the fractional-order constitutive parameters has an appreciable influence on the electrical potential generated across the piezoelectric patch. To better demonstrate the effect of nonlocal interactions, we provide the RMS voltage for each of these cases in Table~\ref{tab:RMS voltage patch}. In this case, the definition of the RMS voltage in Eq.~\eqref{eq: rms} is appropriately modified to account for only the nodes at which the piezoelectric patch is present. Note that in this study, we choose $\mathcal{N}=500$ for all choices of fractional-order length scales.

\begin{table}[H]
\begin{center}
    \begin{tabular}{ c c c c c } 
    \hline
    \hline
    $h_l$ & \multicolumn{4}{c}{$V_{rms}$~(in $\mathrm{V})$} \\ 
    \hline
     & $\alpha=1.0$ & $\alpha=0.9$ & $\alpha=0.8$ & $\alpha=0.7$ \\
     \hline
    $L/20$ & 0.9296 & 0.9284 & 0.9254 & 0.9224 \\[1.5ex]   
     $L/10$ & 0.9296 & 0.9241 & 0.9162 & 0.9087 \\[1.5ex]
     $L/5$ & 0.9296 & 0.9185 & 0.9033 & 0.8879 \\     
    \hline
    \hline
    \end{tabular}
    \caption{RMS voltage of a cantilever smart beam for $q_0(x_1)=1~\mathrm{N/m}$.}
    \label{tab:RMS voltage patch}
\end{center}
\end{table}

The RMS voltages induced across the domain of the piezoelectric patch demonstrate a consistent reduction with an increasing degree of nonlocal interactions. This observation is in contrast to the results for a simply supported smart beam with a piezoelectric layer in Table~\ref{tab:RMS voltage layer}. This can be attributed to a consistent softening of \textit{all} the system matrices in Eq.~\eqref{eq:algebraic_gov_eq_sensing}. Therefore, for the current study of a smart cantilever beam with a piezoelectric patch, the effect of softening of mechanical stiffness (over the entire length of the smart beam) outweighs the softening of electrical system matrices (defined only over the length of the piezoelectric patch). The contrasting results for the influence of nonlocal constitutive parameters over the electrical potential in Tables~\ref{tab:RMS voltage layer} and \ref{tab:RMS voltage patch} points to interesting possibilities in tuning nonlocal interactions for achieving desired electro-mechanical, and thereby any general multiphysics, coupling.

The effect of nonlocal interactions is more pronounced over the electrical potential induced over the piezoelectric layer or patch as seen in Figs.~\ref{fig:Sensing_simply h_l and alpha} and \ref{fig:Sensing_CF h_l and alpha}. This is particularly true for the electrical potential induced over the piezoelectric patch on a cantilever beam. To explain this, it is important to reiterate that the electrical potential induced within the Euler-Bernoulli smart beam through direct piezoelectric coupling is directly proportional to the axial normal strain. This relationship is analogous to that of local piezoelectric constitutive models \cite{krommer2001correction}. Therefore, an increment in the fractional-order axial normal strain due to nonlocal interactions results in a simultaneous increase in the electrical potential induced at the top surface of the piezoelectric layer/patch. Recall that the effect of nonlocal interactions is more pronounced on the strains when compared with displacements\cite{patnaik2020ritz}. This is because of the differ-integral definition for strain-displacement relations that renders the strain at a point to depend on the deformations at all points within the horizon of nonlocal influence (see Eq.~\eqref{eq:RC_def}). This explains a greater difference between local and nonlocal results for electrical potential induced within the piezoelectric layer/patch (proportional to strain) when compared to mechanical displacement presented earlier. Moreover, the effect of truncating the length scales is significant on the nonlocal strain, and therefore, the electrical potential produced in the piezoelectric patch in Fig.~\ref{fig:Sensing_CF h_l and alpha} presents a greater difference between local and nonlocal results, when compared to the piezoelectric layer in Fig.~\ref{fig:Sensing_simply h_l and alpha}.

\section{Conclusions} 
In this study, we develop a constitutive model for nonlocal piezoelectricity employing fractional-order definitions for mechanical strain and electrical field variables. The integro-differential definitions for fractional-order derivatives capture long-range interactions over the elastic and electrical fields. Additionally, multiphysics coupling within the nonlocal solid via piezoelectricity allows the electrical potential at a point to be influenced by mechanical displacement over a horizon of nonlocal influence, and vice versa. To better illustrate the potential of such a coupling between nonlocal field variables, an example of an unimorph smart beam with the piezoelectric patch is considered. In this example, nonlocal effects are studied for two possible cases of direct and converse piezoelectric effects. Modeling the smart beam as following Euler-Bernoulli beam displacement theory, governing fractional-order differential equations are derived for each case mentioned above. Unlike existing integer-order theories for nonlocal piezoelectricity, the fractional-order approach adopted here ensures consistent and well-posed constitutive relations with a unique solution. Thereafter, a numerical solver is developed for these fractional-order differential equations. A series of parametric studies are conducted for different configurations of nonlocal smart beams with varying fractional-order constitutive parameters. The effect of nonlocal interactions is primarily towards the softening of corresponding mechanical, electrical and coupled system (stiffness and force) matrices, and present the improvements in electro-mechanical coupling of the smart structure. This observation opens exciting possibilities towards tuning the degree of electro-mechanical coupling by controlling the degree of nonlocal interactions across the domain of the smart beam. Therefore, this study establishes the fundamental framework for designing exciting piezoelectric, or general multiphysics, coupling-based metastructures by leveraging long-range nonlocal interactions.\\  

\noindent
\textbf{Acknowledgements}: S.D. and S.S. acknowledge the financial support from the Science and Engineering Research Board (SERB), India, under the startup research grant program (SRG/2022/000566).

\bibliographystyle{naturemag}
\bibliography{references}
\newpage

\end{document}